\documentclass[12pt]{article}
\renewcommand{\figurename}{Figure}

\usepackage[dvips]{graphicx}
\usepackage{color}

\newcommand{\mc}{\mathcal}
\newcommand{\real}{\mathbb{R}}

\newcommand{\map}[3]{#1: #2 \rightarrow #3}
\newcommand{\transpose}{\mathsf{T}}

\usepackage{amssymb}
\usepackage{amsfonts}
\usepackage{amsmath}
\usepackage[linesnumbered,ruled,vlined]{algorithm2e}
\usepackage{color}
\usepackage{float}
\usepackage{placeins}

\usepackage{caption}
\usepackage{subcaption}
\usepackage[font=small,skip=0pt]{caption}
\usepackage{graphicx}

\newcounter{lastnote}

\usepackage[labelsep=endash]{caption}

\usepackage{scicite}

\usepackage{times}
\usepackage{graphicx}
\usepackage[hidelinks]{hyperref} 

	\title{Cognitive Control in the Controllable Connectome} 
	
	\author
	{John D. Medaglia$^{1}$, Shi Gu$^{2,3}$, Fabio Pasqualetti$^{4}$, Rebecca L. Ashare$^{5}$,\\ Caryn Lerman$^{5}$, Joseph Kable$^{1}$, Danielle S. Bassett$^{2,3}$\\
		\\
		\normalsize{$^{1}$Department of Psychology, University of Pennsylvania}\\
		\normalsize{Philadelphia, PA, 19104, USA}\\
		\normalsize{$^{2}$Department of Bioengineering, University of Pennsylvania}\\
		\normalsize{Philadelphia, PA, 19104, USA}\\
		\normalsize{$^{3}$Department of Electrical \& Systems Engineering, University of Pennsylvania}\\
		\normalsize{Philadelphia, PA, 19104, USA}\\
		\normalsize{$^{4}$Department of Mechanical Engineering, University of California - Santa Barbara}\\
		\normalsize{Philadelphia, PA, 19104, USA}\\
		\normalsize{$^{5}$Department of Psychiatry, University of Pennsylvania}\\
		\normalsize{Philadelphia, PA, 19104, USA}\\
		\\
		\normalsize{$^\ast$ To whom correspondence should be addressed; E-mail:  dsb@seas.upenn.edu.}
	}
	
\begin{document}
			\maketitle
			\newpage
		\begin{abstract}
			Cognition is supported by neurophysiological processes that occur both in local anatomical neighborhoods and in distributed large-scale circuits. Recent evidence from network control theory suggests that white matter pathways linking large-scale brain regions provide a critical substrate constraining the ability of single areas to affect control on those processes. Yet, no direct evidence exists for a relationship between brain network controllability and cognitive control performance. Here, we address this gap by constructing structural brain networks from diffusion tensor imaging data acquired in 125 healthy adult individuals. We define a simplified model of brain dynamics and simulate network control to quantify modal and boundary controllability, which together describe complementary features of a region's theoretically predicted preference to drive the brain into different cognitive states. We observe that individual differences in these control features derived from structural connectivity are significantly correlated with individual differences in cognitive control performance, as measured by a continuous performance attention test, a color/shape switching task, the Stroop inhibition task, and a spatial n-back working memory task. Indeed, control hubs like anterior cingulate are distinguished from default mode and frontal association areas in terms of the relationship between their control properties and individual differences in cognitive function. These results provide the first empirical evidence that network control forms a fundamental mechanism of cognitive control.

		\end{abstract}
		\newpage
Cognitive control refers to the ability to adaptively vary information processing and subsequent behavior to achieve specific goals. This ability enables humans to link information to solve problems, inhibit inappropriate behavioral responses, switch among tasks, and actively update, select, and maintain behaviorally relevant information.  Failures in cognitive control are ubiquitously observed in neuropsychiatric conditions and predict long-term outcomes ranging from school performance \cite{Normandeau1998,Diamond2007} to career achievement \cite{Diamond2007}.   

Concerted efforts in cognitive neuroscience have demonstrated that these functions have specific anatomical bases in the brain, including a distributed set of areas in frontal, medial, and parietal cortices \cite{Cole2013,Power2013,Voytek2015}. Yet, how these regions work in concert with one another remains far from understood \cite{Medaglia2015}. One emerging perspective is that cognitive control is driven by dynamic interactions between large-scale neural circuits or networks \cite{Cocchi2013,Gu2015,Mattar2015}. These interactions can take the form of coordination or competition \cite{Cocchi2013}, and can change over time to enable behavioral outputs such as button presses and spoken words. Indeed, the network-level processes of cognitive control may facilitate the flexible management of large-scale brain activity in support of successful performance in a variety of effortful tasks \cite{Cole2013,Braun2015,Bassett2015}. 
		
While cognitive control is thought of as a network-level process, putative mechanisms of this process have not been identified. Recent theoretical work posits that network control theory - a nascent field of engineering - offers exactly such a fundamental mechanism \cite{Gu2015}. Network control theory is the study of how to design control strategies for networked systems \cite{Ruths2014}, in which a set of nodes are connected by edges and in which the activity of nodes can be modeled using simplified linear dynamics. These tools predict that different nodes serve dissociable control roles defined by their location on the network \cite{Pasqualetti2014}. In the context of the brain, such a mechanism would suggest that brain regions are predisposed to drive or modulate neurophysiological dynamics in a manner consistent with their specific topological signatures of white matter connectivity \cite{Gu2015,betzel2016optimally}. 

Here we test whether network control is a putative mechanism of cognitive control by asking whether the theoretically predicted control features of brain regions are related to cognitive performance on so-called cognitive control tasks. We restrict our attention to two distinct network control features known as modal controllability and boundary controllability. Intuitively, modal controllability describes the ability of a node (here, a brain region) to drive a network into difficult-to-reach states on an energy landscape, thus providing a means for switching between disparate tasks. Based on prior work locating modal controllers in fronto-parietal cognitive control systems \cite{Gu2015}, we hypothesize that individual differences in the modal controllability of these areas will be associated with individual differences in performance on demanding cognitive control tasks such as those with high working memory or inhibition demands, or switching costs. To complement these intuitions, we also study boundary controllability, which describes the ability of a node to steer the system into states where modules are either coupled or decoupled, thus providing a means for coordination and competition. Based on prior work locating boundary controllers in attention systems \cite{Gu2015}, we hypothesize that individual differences in the boundary controllability of these areas will be associated with individual differences in the performance of attention-demanding tasks.

	\begin{figure}[htp]
		\centerline{\includegraphics[width=3.5in]{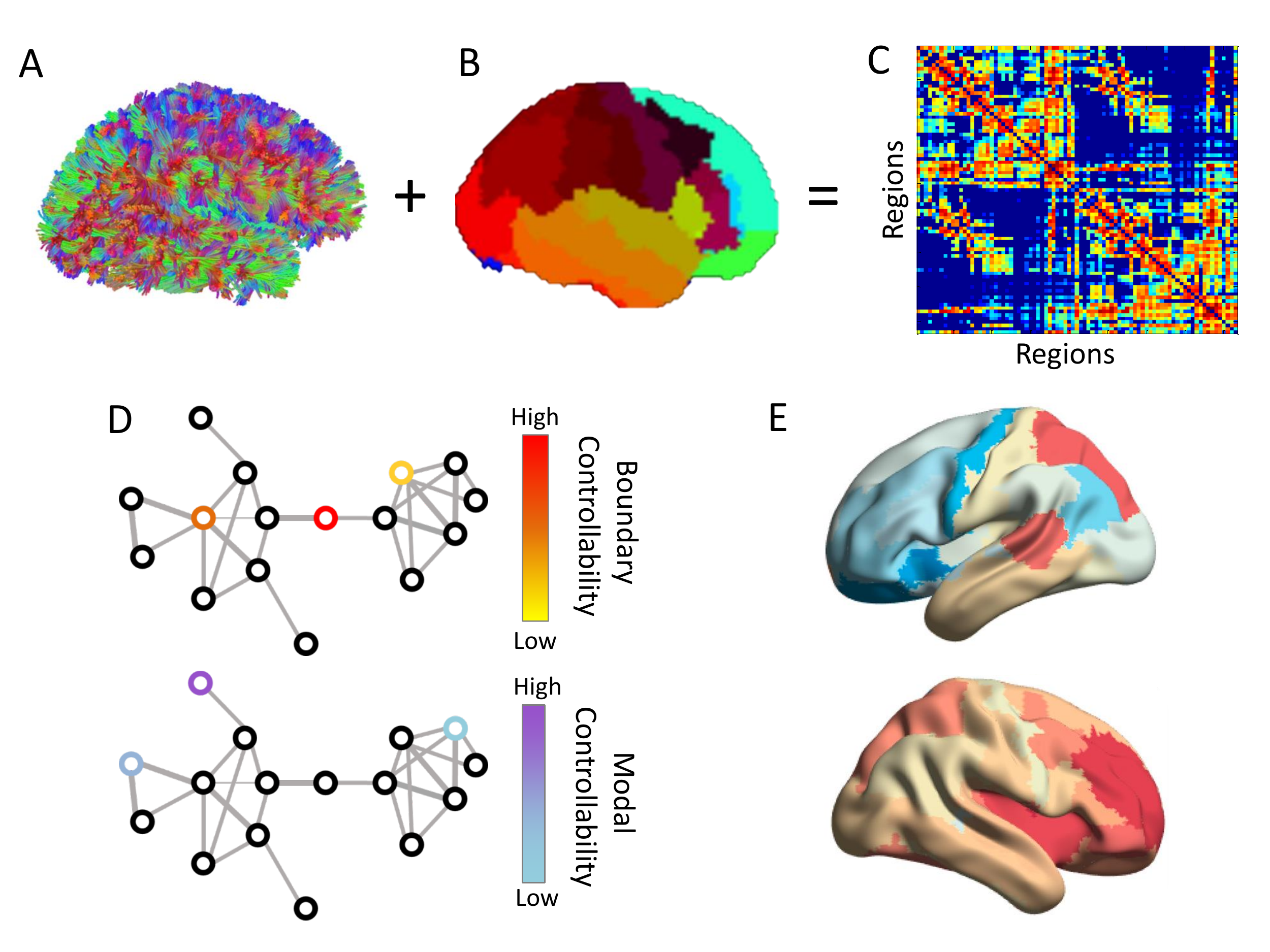}}	
		
		\setlength{\belowcaptionskip}{-12pt}
		\caption{\textbf{Overview of Methods} \emph{(A)} We perform diffusion tractography for each subject, and \emph{(B)} apply a whole-brain parcellation to identify anatomical divisions. \emph{(C)} We construct an adjacency matrix that represents the number of streamlines between pairs of regions. \emph{(D)} We define a simplified model of brain dynamics and simulate network control to quantify modal and boundary controllability for each node (brain region) in the network for each individual. \emph{(E)} We map the strength of correlation between regional controllability and cognitive performance across individuals.}\label{methods}\end{figure}

To test these hypotheses, we construct structural brain networks from diffusion tensor imaging (DTI) data acquired in each of 125 healthy adult subjects (Fig.~\ref{methods}A). Each network contains 83 brain regions defined by the Lausanne anatomical parcellation (Fig.~\ref{methods}B), and each pair of regions is connected by an edge weighted by the number of streamlines linking those regions (Fig.~\ref{methods}C). We define a simplified model of brain dynamics and simulate network control to quantify modal and boundary controllability (Fig.~\ref{methods}D). We reveal intersubject variability in node controllability that is associated with different cognitive control functions \cite{miyake2000unity}, either being advantageous or disadvantageous for cognitive control performance (Fig.~\ref{methods}E). Performance is measured by the number of true positives on the continuous performance test \cite{kurtz2001comparison}, color/shape median switch cost (reaction time) \cite{miyake2004inner}, median Stroop effect (reaction time) \cite{stroop1935studies}, and a visuo-spatial n-back \cite{green2005muscarinic}. The continuous performance test measures sustained attention or vigilance. The color/shape task median switch cost measures an individual's ability to switch between task sets efficiently. The Stroop effect measures an individual's ability to inhibit prepotent responses to semantic information. The visuo-spatial n-back measures the ability to hold and update a limited amount of information in mind. By examining the relationship between interindividual variability in controllability and distinct cognitive control functions, we establish a bridge between a neuroscientist's notion of cognitive control and an engineer's notion of network control, suggesting that the latter forms a mechanism for the former.

\section*{Results}
\subsection*{Individual Variability in Cognitive Measures}
\begin{figure}[htp]
	\centerline{\includegraphics[width=3.5in]{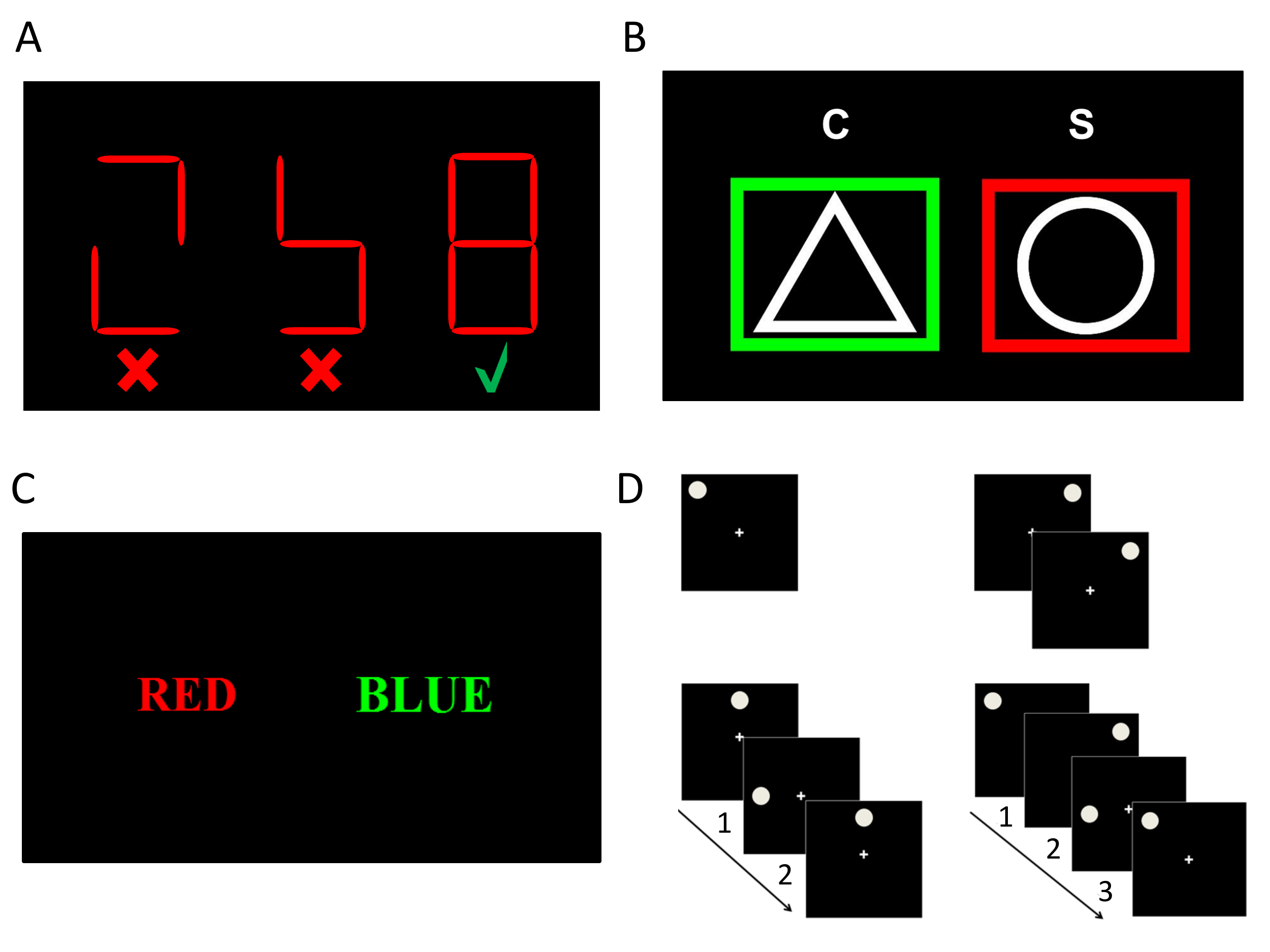}}
	\caption{\textbf{Task Stimuli.} All stimuli were presented on a computer screen on a black background. \emph{(A)} Continuous performance test: participants are asked to respond only to complete numbers. \emph{(B)} Color/shape task. Left: ``C" cues the participant to respond to the color of the bars. Right: ``S" cues participants to respond to the shape. Switches between these conditions produce a shift cost. \emph{(C)} Stroop task: left, a congruent trial; right, an incongruent trial requiring inhibition of the word form. \emph{(D)} Four difficulty levels of the visual-spatial n-back. Top left: an example single stimulus. Remaining panels: participants were asked to respond when the current stimulus matched a stimulus 1, 2, or 3-back from the current stimulus.}\label{Tasks}\end{figure}

All cognitive measures were collected as part of a testing battery designed to assess cognitive control and decision making (see Methods for details; see Fig.~\ref{Tasks}). The four tests we study were selected based on (1) their statistical independence from one another, (2) their use in the measurement of distinct cognitive control processes, and (3) their sensitivity to disease diagnoses. We observed that these four tests contained little shared variance with one another in explaining individual differences in cognitive control performance ($R^{2} = 0.001:0.08$). These results indicate that the four measures represent unique information that can assist in distinguishing the cognitive control abilities of one person from those of another.

\subsection*{Cognitive Control Performance and Network Controllability}

We next ask whether individual differences in cognitive control performance (as measured by these tasks) are related to individual differences in white matter architecture, the substrate of an engineer's notion of structural network control. To address this question, we calculate the Spearman correlation coefficient between regional controllability values and performance. To determine statistical significance, we apply a spatial test to determine non-trivial structure in the anatomical distributions of these correlations (see Methods). Briefly, we determine the statistical similarity (Pearson correlation) between the inverse Euclidean distance between node centroid pairs and the mean absolute value of cognition-control correlations for the same node pair. A positive correlation demonstrates that the spatial proximity of nodes is related to the strength of their controllability relationship to cognition. 

We report results for each of the four cognitive control performance measures, as well as for each of the two regional controllability statistics (modal and boundary). Across all eight analyses, we observed that each map demonstrated highly significant spatial organization in the relationships between individual differences in controllability and cognitive performance. For modal controllability, we observed the following relationships: CPT: $r = 0.51$, $df = 6888$, $p = 4.19 \times 10^{-230}$; Color/Shape: $r = 0.38$, $df = 6888$, $p = 1.25 \times 10^{-120}$; Stroop: $r = 0.55$, $df = 6888$, $p = 1.35 \times 10^{-295}$; N-Back: $r = 0.57$, $df = 6888$, $p = 1.15 \times 10^{-302}$. For boundary controllability, we observed the following relationships: CPT: $r = 0.57$, $df = 6888$, $p = 5.71 \times 10^{-306}$; Color/Shape: $r = 0.46$, $df = 6888$, $p = 2.19 \times 10^{-181}$; Stroop: $r = 0.43$, $df = 6888$, $p = 7.59 \times 10^{-155}$; N-Back: $r = 0.35$, $df = 6888$, $p = 1.15 \times 10^{-302}$. These results demonstrate that brain regions whose controllability values are associated with individual differences in cognitive control performance display non-trivial spatial clustering.   

Importantly, these maps reveal differential relationships between controllability and cognitive control: increases in controllability in some regions are advantageous for cognitive control performance, whereas increases in others are disadvantageous. In the following sections, we examine the maps for each cognitive control performance measure separately, report significant regional predictors of performance, and interpret their anatomical locations. 
	
\subsection*{Regional Controllers in the Continuous Performance Task}

\begin{figure}[h!]
	\centerline{\includegraphics[width=3.5in]{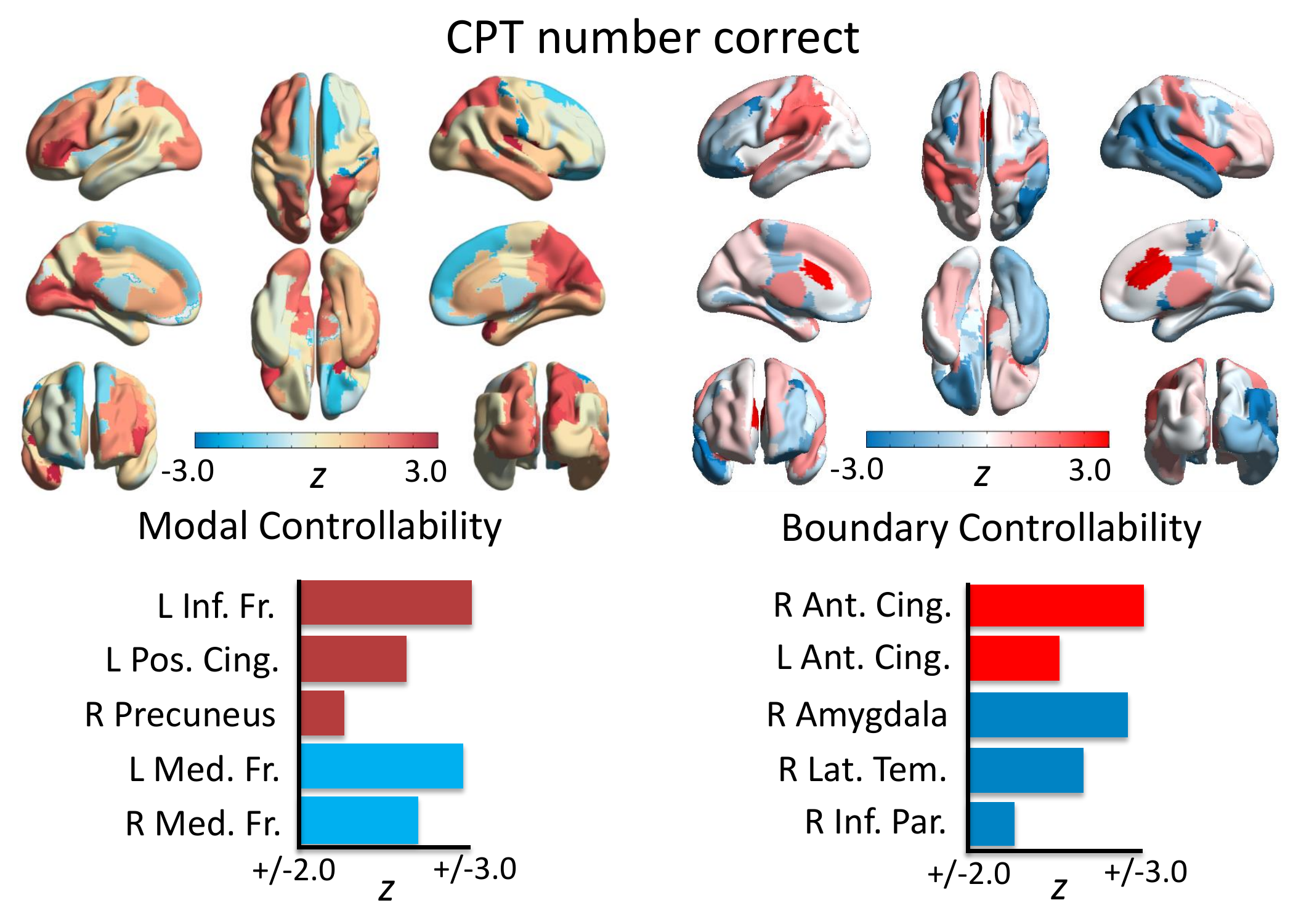}}
	\caption{\textbf{Correlations between controllability and true positives on the Continuous Performance Test.} Results for the association between individual differences in modal and boundary controllability of structural brain networks and true positives on the continuous performance test. Both maps represent the spatial distributions of Fisher's $r$ to $z$ transformed values of the relationship between controllability and cognition. Warmer colors indicate a positive association between controllability and cognition, whereas cooler colors indicate a negative association between controllability and cognition. R = right, L = left, Inf = inferior, Ant. = anterior, Pos. = posterior, Cing. = Cingulate, Med. = medial, Lat. = lateral, Fr. = frontal, Tem. = temporal, Par. = parietal.}\label{CPT}
\end{figure}

We begin by examining the anatomical distribution of correlations between controllability values estimated from structural brain networks and the number of true positives on the continuous performance test (Fig.~\ref{CPT}). We interpret regions for which $r$-to-$z$ transformed values of correlation surpass $+/- 2.0$. We observe that modal controllability in the left inferior frontal gyrus, posterior cingulate, and right precuneus are most positively associated with performance, whereas controllability in the bilateral medial frontal gyri were most negatively associated with performance. Boundary controllability in the bilateral caudal anterior cingulate are advantageous for performance, whereas increasing strength in the right inferior parietal lobe, lateral temporal lobe and amygdala are relatively disadvantageous. 

The anatomical loci of these control-cognition relationships are particularly interesting in light of the posited neuroanatomical basis of continuous performance task execution. For example, recent functional neuroimaging studies have identified activation in the cingulate cortex and precuneus during continuous performance tests \cite{ogg2008neural}. More recently, data has further supported an association between disrupted posterior cingulate function and arousal states, including states of internal or externally focused attention, and states modulating breadth of mental focus \cite{leech2014role}. The precuneus is typically suppressed during attention processing, and its deactivation is associated with stability of attention \cite{sali2016spontaneous}. Moreover, sustained attention is also associated with activation in anterior cingulate cortex, which is presumed to contribute to associated error monitoring \cite{botvinick2004conflict}. More generally, the inferior parietal lobe and medial frontal gyrus have been associated with attention shifts \cite{nagahama1999transient}, and reduced activity in the lateral temporal lobe has been associated with reduced attention switching and inhibition performance \cite{smith2006task}. Furthermore, the ventral nucleus of the amygdala has been associated with attentional function in fear conditioning \cite{holland1999amygdala}.
	
In light of this body of work in cognitive neuroscience, our results offer a fresh perspective on the potential structural mechanisms subserving these observations. Specifically, our findings suggest a tradeoff between advantageous and disadvantageous regional control roles: if regions that contribute to attention switching and emotional reactivity have an increased role in controlling the brain, maintaining sustained attention is more difficult. Conversely, stronger controllability situated in regions associated with sustaining attention leads to more stable attention in the continuous performance test. Sustained attention contributes to many domains of cognitive function, and is ubiquitously effected in neuropsychiatric disorders \cite{shanmugan2016common}. Here, we find that increasing strength in the role of emotion processing and attention switching regions in driving brain wide dynamics may be one basis for these observed deficits. 
	
\subsection*{Regional Controllers in the Color Shape Task}
	
\begin{figure}[h!]
		\centerline{\includegraphics[width=3.5in]{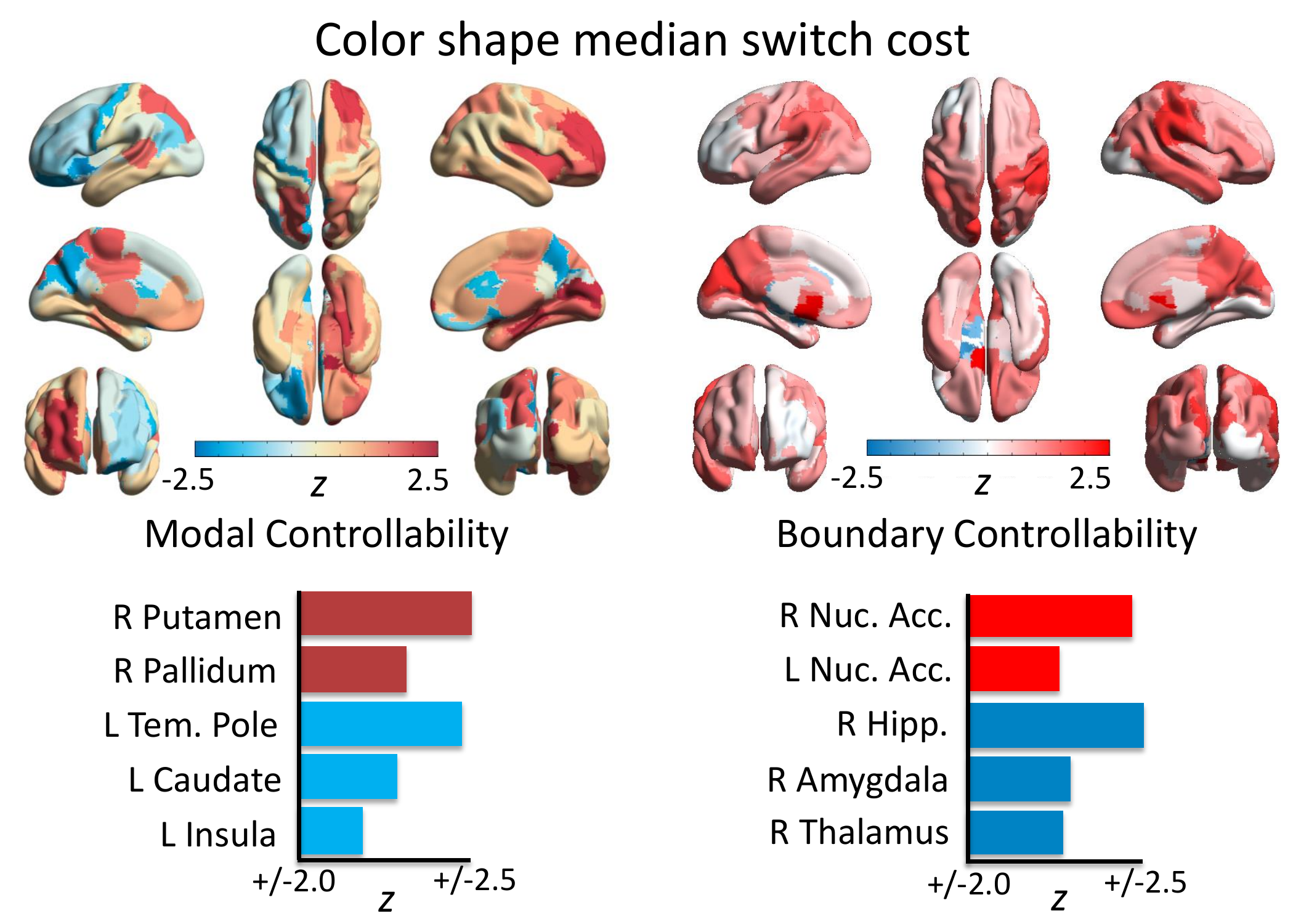}}
		\caption{\textbf{Correlations between controllability and median switch cost on the Color Shape task.} Results for the association between individual differences in modal and boundary controllability of structural brain networks and median switch cost on the color shape task. Both maps represent the spatial distributions of Fisher's $r$ to $z$ transformed values of the relationship between controllability and cognition. Warmer colors indicate a positive association between controllability and cognition, whereas cooler colors indicate a negative association between controllability and cognition. R = right, L = left, Nuc. Acc. = nucleus accumbens, Hipp. = hippocampus, Ant. = anterior, Tem. = temporal} \label{ColorShape}
	\end{figure}

Next, we examine the anatomical distribution of correlations between controllability values estimated from structural brain networks and the median switch costs on the color shape task (Fig.~\ref{ColorShape}). Again, we interpret regions for which $r$-to-$z$ transformed values of correlation surpass $+/- 2.0$. We observe that modal controllability in the right putamen and pallidum is positively associated with performance (reduced switch costs), whereas modal controllability in the left temporal pole, caudate, and insula is negatively associated with performance (enhanced switch costs). Boundary controllability in the bilateral nucleus accumbens is positively associated with performance, whereas boundary controllability in the right hippocampus, amygdala, and thalamus is negatively associated with performance. 

The posited neuroanatomical basis of task switching paradigms traditionally place a special emphasis on the role of prefrontal regions \cite{Cole2013,Power2013}. However, a broader view acknowledges that subcortical systems heavily interact with fronto-parietal areas supporting goal maintenance and switching \cite{isoda2007switching,hikosaka2010switching,floresco2006dissociable}. Moreover, subcortical systems are critical for parsing and consolidating task representations during multitasking \cite{garner2015training}, and focal lesions in the basal ganglia result in errors when trial responses require the application of the correct task rule \cite{aron2007converging}. Indeed, more generally, the basal ganglia are anatomically well-positioned to regulate information flow among multiple cortical regions, contributing highly regular computational functions in parallel-projecting segregated loops \cite{parent1995functional}. 

Our results complement these prior efforts by suggesting that strengthening in the role of the putamen and pallidum in driving the brain into difficult to reach states is associated with reduced switch costs, whereas strengthening in the role of the nucleus accumbens into integrated and segregated states is associated with reduced switch costs. In light of these findings, it is intuitively plausible that task-switching deficits in classically subcortical neurodegenerative diseases such as Parkinson's disease may be attributable to disruption of the basal ganglia's role in mediating trajectories that switch cognitive sets \cite{cools2001mechanisms}.  Interestingly, in contrast to the supportive role of the putamen and pallidum, we find that high modal controllability in the caudate is disadvantageous for task switching, as is high boundary controllability in the hippocampus, amygdala, and thalamus. These findings indicate that as regions that are relatively specialized for long term memory storage and consolidation, reinforcement and reward learning, sensory processing and relays, and planned action in unpredictable environments become more dominant in the control hierarchy, fluid cognitive transitions may become more difficult. It is conceivable that relative increases in their role regulating intermodular communication competes with the role of the basal ganglia in regulating cortical dynamics during cognitive transitions. This may potentially explain observations of deficits during task switching in anxiety \cite{derakshan2009effects}, which is associated with altered function and structure in the hippocampus, thalamus, and amygdala \cite{gross2004developmental}.

\subsection*{Regional Controllers in the Stroop Task}

\begin{figure}[h!]
	\centerline{\includegraphics[width=3.5in]{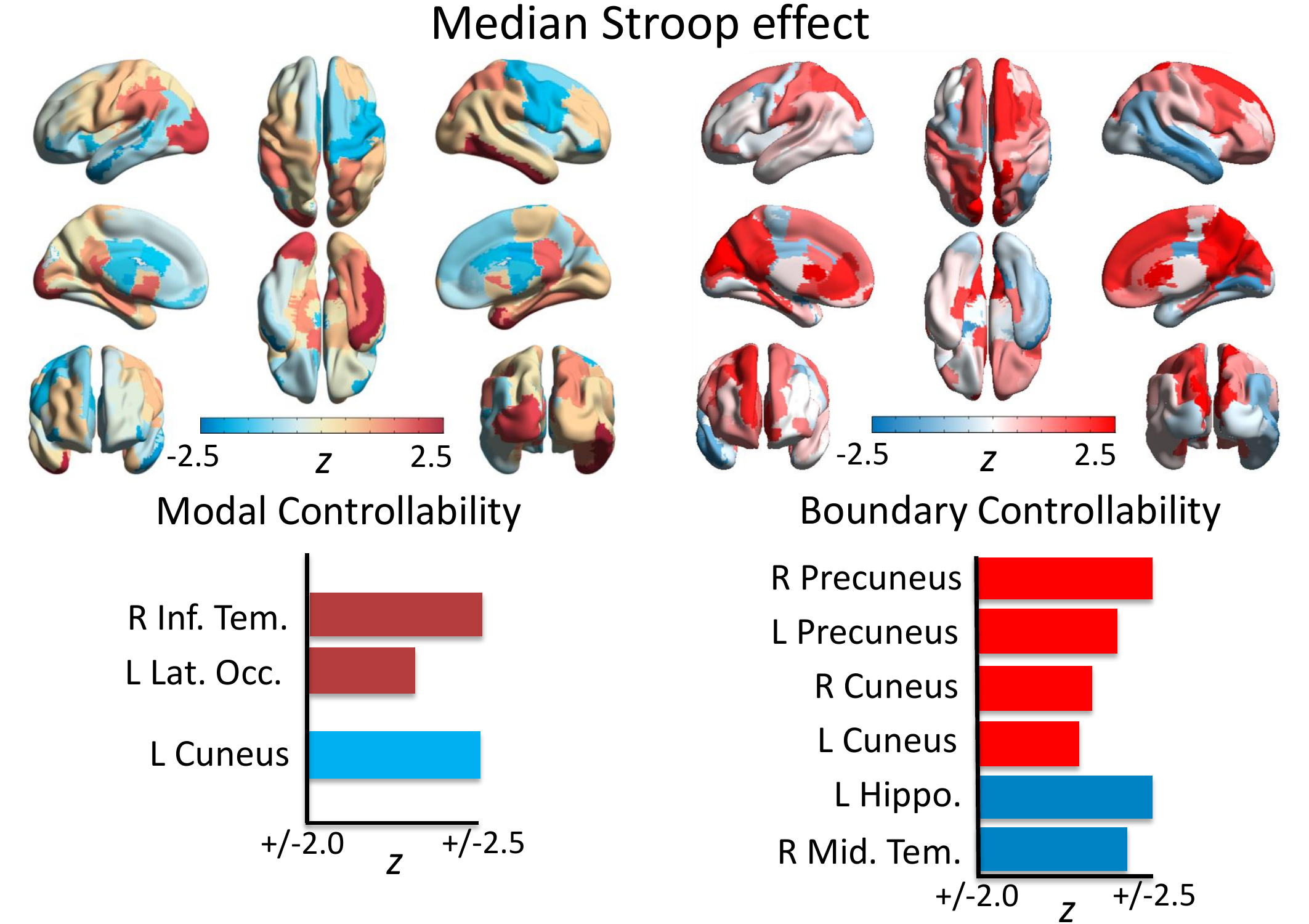}}
	\caption{\textbf{Correlations between controllability and the median response time on the Stroop task.} Results for the association between individual differences in modal and boundary controllability of structural brain networks and the median reaction time on the Stroop task. Both maps represent the spatial distributions of Fisher's $r$ to $z$ transformed values of the relationship between controllability and cognition. Warmer colors indicate a positive association between controllability and cognition, whereas cooler colors indicate a negative association between controllability and cognition. R = right, L = left, Hipp. = hippocampus, Mid. = middle, Lat. = lateral, Tem. = temporal, Occ = occipital.}\label{Stroop}
	\end{figure}

Next, we examine the anatomical distribution of correlations between controllability values estimated from structural brain networks and the median response time on the Stroop task (Fig.~\ref{Stroop}). We observe that modal controllability in the right inferior temporal and left lateral occipital lobe is positively associated with performance, whereas modal controllability in the left cuneus is negatively associated with performance. Boundary controllability in the bilateral precuneus and cuneus is positively associated with performance, whereas boundary controllability in the left hippocampus and right middle temporal gyrus is negatively associated with performance. 
	
The anatomical loci of these control-cognition relationships are directly relevant to Stroop activations, as well as to regions known to process different facets of the Stroop stimuli. For example, the left lateral occipital cortex supports right hemifield processing, is selectively activated in processing letter strings relative to face and texture processing \cite{puce1996differential}, and is involved in accessing orthographic whole-word representations \cite{ludersdorfer2015accessing}. The right inferior temporal gyrus participates in the ventral stream pathway, integrating color and shape information during object recognition and recall \cite{creem2001defining}. The current results suggest that the ease in suppressing the prepotent effects of lexical information in the Stroop task is partially attributable to the ability of these regions to propel the brain into difficult-to-reach states. 
	
Prior work using fMRI to investigate Stroop task performance have demonstrated the recruitment of the precuneus in the presence of many or few distractors \cite{banich2000fmri}. The current results suggest that the dynamic role of the precuneus and its support of performance depends on its underlying structural profile: increased involvement in regulating intermodule processing, potentially facilitating information suppression early after visual stimulus representation, reduces the Stroop effect, whereas an increased role in driving the brain into hard to reach states exaggerates the Stroop effect.

\subsection*{Regional Controllers in the N-Back Working Memory Task}

\begin{figure}[h!]				\centerline{\includegraphics[width=3.5in]{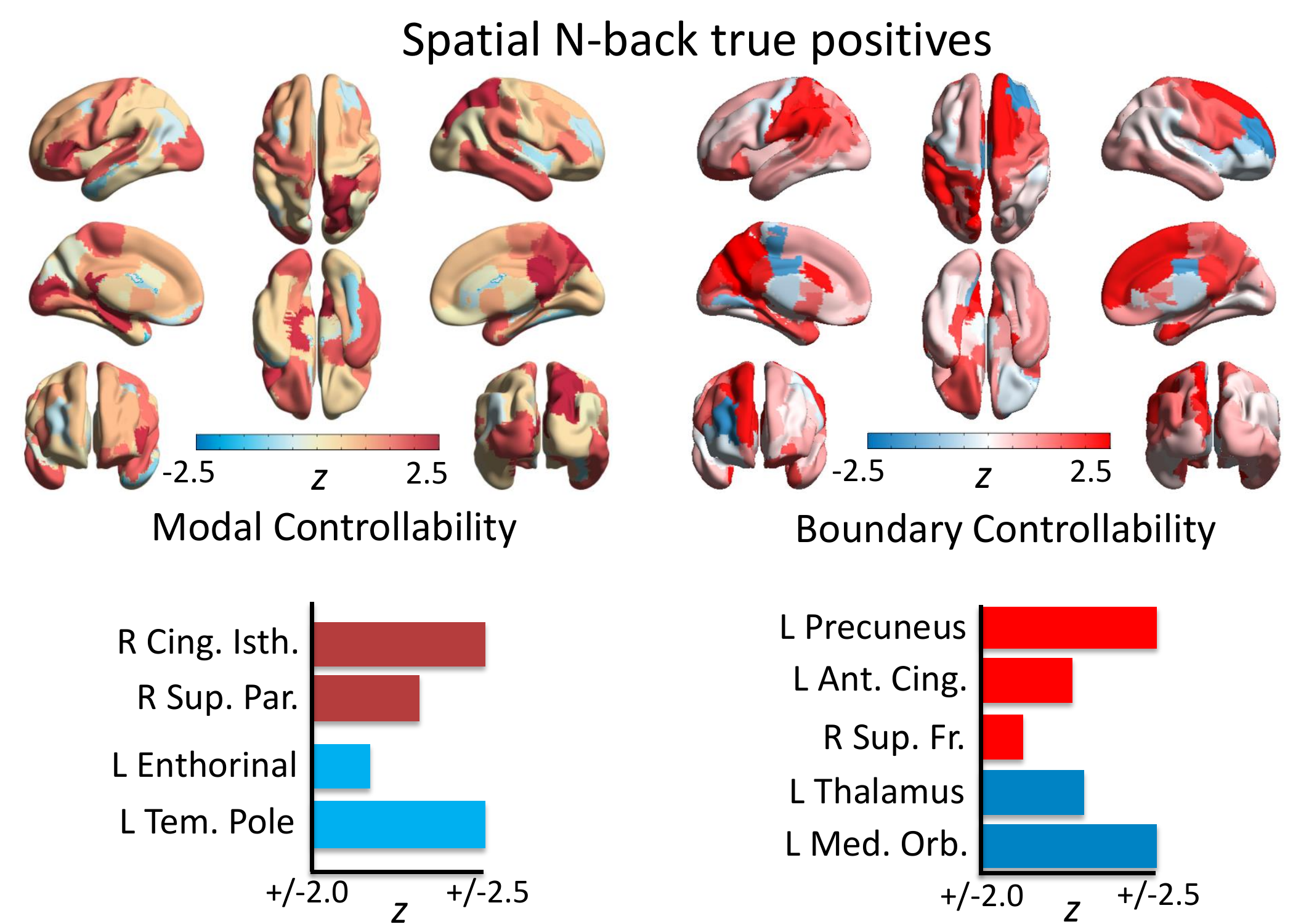}}
	\caption{\textbf{Correlations between controllability and true positives on the spatial n-back task.} Results for the association between individual differences in modal and boundary controllability of structural brain networks and the number of true positives on the spatial n-back task. Both maps represent the spatial distributions of Fisher's $r$ to $z$ transformed values of the relationship between controllability and cognition. Warmer colors indicate a positive association between controllability and cognition, whereas cooler colors indicate a negative association between controllability and cognition. R = right, L = left, Ant. = anterior, Cing. = cingulate, Isth. = isthmus, Med. = medial, Orb. = orbitofrontal, Sup. = superior, Fr. = frontal, Tem. = temporal, Par. = parietal.}\label{NBack}
\end{figure}

Finally, we examine the anatomical distribution of correlations between controllability values estimated from structural brain networks and true positives on a spatial n-back task (Fig.~\ref{NBack}). We observe that modal controllability in the right cingulate isthmus and superior parietal lobe is positively associated with performance, whereas modal controllability in the left entorhinal and temporal pole is negatively associated with performance. Boundary controllability in the left precuneus and anterior cingulate cortex and right superior frontal gyrus is positively associated with performance, whereas boundary controllability in the left thalamus and medial orbitofrontal gyrus is negatively associated with performance. 
	
Again, the anatomical loci of these control-cognition relationships are directly relevant to n-back memory performance and consistent with previous functional neuroimaging findings. For example, the superior parietal lobe contributes to the maintenance and manipulation of information in working memory \cite{koenigs2009superior}, and contributes to spatial attention processing \cite{vandenberghe2012spatial}. The cingulate isthmus plays a more general role, being anatomically positioned to regulate activity in the default mode network and the frontoparietal network recruited for memory performance \cite{braga2013echoes}.  Similar to the CPT, increasing boundary controllability in the left anterior cingulate cortex was advantageous for performance, potentially enabling intermodular communication of error-related signals. Finally, the superior frontal gyrus supports working memory and attention \cite{petrides2000role}, and participates in the frontoparietal control system responsible for moment-to-moment response monitoring and set maintenance \cite{Power2011}. Increases in this region's modal control role may allow the brain to better maintain the difficult states necessary to facilitate monitoring among distractors.
	
Interestingly, in contrast to the supportive role of these areas, high control values in entorhinal cortex, temporal pole, and orbitofrontal gyrus is be detrimental to performance.  To understand these results, we recall that the left entorhinal cortex and temporal pole contribute to familiarity for words \cite{brandt2016selective} and binding perceptual and emotional information \cite{olson2007enigmatic}, respectively, and the orbitofrontal gyrus serves a key role in decision making, especially in the context of emotional and reward processing \cite{bechara2000emotion}. Thus, our results suggest that dynamic brain trajectories driven more prominently by computations specialized for lexical recognition, emotional, and sensory processes may interfere with processes mediated through the cingulate and parietal systems. 
	 
It is interesting to consider these anatomical distributions of control-cognition relationships in the context of neuropsychiatric disorders. Specifically, in borderline personality disorder, a syndrome with substantial emotional processing and working memory deficits, orbitofrontal cortex dysfunction is associated with increased impulsive errors on attention demanding and working memory tasks \cite{berlin2005borderline} and altered temporal pole activity has been associated with social cognition and empathy dysfunction \cite{dziobek2011neuronal}. An interesting future direction would be to test for alterations in structural controllability in these regions in clinical syndromes that present with emotion regulation and cognitive control dysfunction.

\section*{Discussion}
   
   In this paper, we address the hypothesis that network control is a fundamental mechanism of cognitive control \cite{Gu2015}. We explicitly test this hypothesis by determining whether individual differences in putative control roles of brain regions are related to individual differences in performance on so-called cognitive control tasks. We construct structural brain networks from diffusion tensor imaging data acquired in 125 healthy adult individuals. We define a simplified model of brain dynamics and simulate network control to quantify modal and boundary controllability, which together describe complementary features of a region's theoretically predicted preference to drive the brain into different cognitive states. We find that brain regions whose controllability values are associated with individual differences in cognitive control performance display non-trivial spatial clustering. Moreover, we find that distinct areas of the brain display controllability values that are strongly correlated with performance in a continuous performance test, a color-shape switching task, a Stroop task, and a n-back working memory task across individuals. Critically, these control-cognition relationships were two-sided: increases in controllability in some regions are advantageous for cognitive control performance, whereas increases in others are disadvantageous.  These results provide the first evidence to support the hypothesis that human brains use network control to affect cognitive control.
   
   \subsection*{Network Control Theory: a Framework to Study Cognition}\
   Network control theory offers a novel framework for the study of human cognition. Unlike graph theory and its associated statistics \cite{Bullmore2009,Rubinov2010,Bullmore2011}, network control theory is built on a model of brain dynamics, which describes patterns of inter-region activity propagated along the wires of an underlying structural network \cite{Pasqualetti2014}.  From such a model, one can mathematically derive theoretical predictions about how the structural network organization of the brain impacts on brain function \cite{Gu2015}. These tools can be tuned to study distinct types of control strategies.  Here we focus on (i) modal controllability, which describes the role of a region in driving the brain into difficult-to-reach states, and (ii) boundary controllability, which describes the role of a region in driving the brain into states of network segregation or integration. This approach represents a substantial but complementary departure from traditional analyses of functional and diffusion imaging data at the voxel, tract, and graph levels.  Instead, network controllability analysis can be used to better understand the role of each brain region in regulating whole-brain network function based explicitly on its structural fingerprint and a working model of node dynamics.  
   
   \subsection*{Relationship to Prior Theoretical Predictions}\
   Prior work identified a spatial dissociation between different types of brain network controllers \cite{Gu2015}: modal controllers are prevalent in fronto-parietal cognitive control regions and boundary controllers are prevalent in dorsal and ventral attention regions. Based on this work, we anticipated that high levels of boundary controllability in anterior cingulate and opercular systems would be advantageous for attention. Further, we anticipated that high levels of modal controllability in frontoparietal systems would be associated with better working memory performance and task switching. Whereas we found strong support for the former hypothesis in the continuous performance test, we found weaker support for the latter hypothesis. Instead, we observed that individual differences in modal and boundary controllability in subcortical, default mode and lower association regions \cite{Sepulcre2014} were observed to positively relate to individual differences in cognitive control performance. These results suggest that while network controllers may be spatially clustered in certain brain subsystems \cite{Gu2015}, interindividual variability in regions that interact with frontoparietal systems through multiple pathways \cite{Hellyer2014} may enhance brain-wide dynamics to support cognitive control functions.
   
   \subsection*{Integrative Insights across Cognitive Control Tasks}\
   	
   In each cognitive task, we identified regions whose variation in controllability correlated with individual differences in performance. In some cases, these regions were consistent with those identified in activation-based studies as pertinent to the task, and in other cases, these regions were novel anatomical loci whose putative functions were supportive of the task. As an example of the former: the contributions of the anterior cingulate cortex in the CPT and N-Back tasks are consistent with classically defined roles for these regions in cognitive neuroscience. Our work extends this body of knowledge by suggesting that this region also plays a role in regulating the integration and segregation of network-wide dynamics, supporting its error monitoring, shifting and updating.  As an example of the latter: the contributions of the basal ganglia in the color/shape task differ from the common emphasis on fronto-parietal and cingulo-opercular systems. Our results suggest that greater controllability in the putamen and pallidum (modal) as well as the nucleus accumbens (boundary) is supportive of better performance, potentially identifying a key role in regulating the brain via pathways involving cortico-basal ganglia-thalamo-cortical loops  \cite{parent1995functional}.
   
   \subsection*{Competing Effects of Controllability on Cognition}\
   	
   	Numerous functional tradeoffs exist in the brain \cite{wang2008functional}. Cognitively, this includes the well-known oppositions between speed and accuracy \cite{chittka2009speed}, between value and time \cite{Pinker1999mind}, and between exploration and exploitation \cite{cohen2007should}. The approach we take in this work offers a novel dimension of dissociable effects on cognition: namely, that for some brain regions higher controllability values are benefitial for cognitive performance, while for others they are detrimental.  While we do not yet know the fundamental mechanisms of these dissociations, it is interesting to speculate based on the theoretical intuitions behind the analytical approach. Specifically, in each task, the increasing controllability of regions may indicate dynamic interference from regions that do not contribute to the primary processing of stimuli or to the execution of network control that supports performance. It is also possible that higher controllability values in some regions may be detrimental to one cognitive function but supportive of another cognitive function, a possibility that could be examined directly in future empirical work. 
   			
   	\subsection*{Linking Cognitive Control and Network Control}\
   		
   	From an engineering perspective, network control is a process in which a system's dynamics are shifted or guided into new trajectories to support specific system goals. Such a process forms a natural description of cognitive control, a transient process putatively deployed to shift or guide the brain into new mental states in support of higher-order cognition. Prior work has reported that different brain regions are theoretically predicted to be effective at different control strategies: (i) regions in fronto-parietal cortex are suggested to be effective as modal controllers, moving the brain into difficult-to-reach states on an energy landscape, and (ii) regions in attentional circuits are suggested to be effective as boundary controllers, moving the brain into states in which cognitive subsystems or modules are either coupled or decoupled from one another \cite{Gu2015}. Our results provide the first empirical evidence that individual differences in structural controllability of single brain regions can explain significant variance in individual differences in cognitive control abilities, as measured by a battery of cognitive tests.  These results support the possibility that network control might form a fundamental mechanism of cognitive control. This perspective complements other computational models of cognitive control \cite{botvinick2014computational} by highlighting the role of white matter architecture on large-scale functional dynamics.

   \subsection*{Methodological Considerations}\
   	
   	Diffusion tensor imaging may undersample some white matter fibers, particularly those linking hemispheres or those that cross paths with other fibers \cite{wedeen2008diffusion}. Future efforts could use diffusion spectrum imaging to improve estimates of structural network architecture.  In addition, the cognitive organization of cognitive control remains contested, and it is possible that behavioral measures other than those studied here might elucidate other aspects of the relationship between cognitive control and network controllability. Furthermore, we note that the inter-measure behavioral correlations here are somewhat lower than those measured in prior factor analyses \cite{Miyake2001}. While this allows us to speak to differences in cognitive control domains of function, future studies should investigate the generalizability of both the inter-measure relationships among cognitive control measures and robustness of associations with network controllability in samples of various socioeconomic backgrounds and ages. Finally, while we focus on modal and boundary controllability as an extension of previous research,  other control strategies may also be relevant for the study of cognitive control in humans \cite{betzel2016optimally,Gu2015,Motter2015}. 
   	
   \subsection*{Conclusion}
   	Our results link classic findings in cognitive neuroscience to the emerging field of network control theory. Moreover, they extend classical views of cognitive control by explicitly acknowledging the role of white matter organization in regulating the dynamic patterns of activity evolving on top of it. Future work could address the potential utility of these conceptual constructs in understanding deficits in cognitive control observed in neurological disease and psychiatric disorders.

\section*{Methods}
Methods and any associated references are available in the online version of the paper.

\subsection*{Subjects}\
Previous analyses involving diffusion tensor imaging tractography have found significant relationships with cognition in samples less than 20 individuals \cite{Kraus2007white}. However, the relationship between white matter network controllability and cognition has not been previously examined. To examine this relationship, we included the largest available sample of individuals with good quality diffusion images to examine our hypotheses. One hundred sixty six individuals completed a testing battery prior to a cognitive training intervention (study funding provided by NCI grant R01-CA170297) during daylight hours. Of these individuals, 145 had DTI images and a T1 acquired as part of a longer neuroimaging protocol. Using a criterion independently validated on the same scanner used to acquire these data, we excluded 17 individuals due to a signal to noise ratio of less than 6.47 \cite{Roalf2016}. We removed a further three individuals for mean relative displacement greater than $3.0mm$ in a separate resting BOLD time series to reduce the likelihood that our results were influenced by subject motion. The final sample included 125 individuals (mean age $= 24.4$, St.D. $=  4.6$, range $= 18 - 34$, $53$ females).  All participants volunteered with written informed consent in accordance with the Institutional Review Board/Human Subjects Committee, University of Pennsylvania.
\newpage
\subsection*{Diffusion Tractography}\

Data acquisition and structural network construction were performed as in previous work \cite{gu2015controllability}. DTI scans sampled 30 directions using a Q5 half-shell acquisition scheme with a maximum $b$-value of 1,000 and an isotropic voxel size of 2.0 mm. We utilized an axial acquisition with the following parameters: repetition time (TR) = 8 s, echo time (TE)= 82 ms, 70 slices, field of view (FoV) (230, 230, 140 mm). 

DTI data were reconstructed in DTI Studio (www.dsi-studio.labsolver.org) using $q$-space diffeomorphic reconstruction (QSDR) \cite{yeh2011estimation}. QSDR first reconstructs diffusion-weighted images in native space and computes the quantitative anisotropy (QA) in each voxel. These QA values are used to warp the brain to a template QA volume in Montreal Neurological Institute (MNI) space using the statistical parametric mapping (SPM) nonlinear registration algorithm. Once in MNI space, spin density functions were again reconstructed with a mean diffusion distance of 1.25 mm using three fiber orientations per voxel. Fiber tracking was performed in DSI Studio with an angular cutoff of 55$^\circ$, step size of 1.0 mm, minimum length of 10 mm, spin density function smoothing of 0.0, maximum length of 400 mm and a QA threshold determined by DWI signal in the colony-stimulating factor. Deterministic fiber tracking using a modified FACT algorithm was performed until 1,000,000 streamlines were reconstructed for each individual.

Anatomical scans were segmented using FreeSurfer \cite{fischl2012freesurfer} and parcellated using the connectome mapping toolkit \cite{cammoun2012mapping}. A parcellation scheme including $n=129$ regions was registered to the B0 volume from each subject's DTI data. The B0 to MNI voxel mapping produced via QSDR was used to map region labels from native space to MNI coordinates. To extend region labels through the grey-white matter interface, the atlas was dilated by 4 mm \cite{cieslak2014local}. Dilation was accomplished by filling non-labelled voxels with the statistical mode of their neighbors' labels. In the event of a tie, one of the modes was arbitrarily selected. Each streamline was labelled according to its terminal region pair. From these data, we constructed a structural connectivity matrix, $\mathbf{A}$ whose element $A_{ij}$  represented the number of streamlines connecting different regions, divided by the sum of volumes for regions $i$ and $j$.

\subsection*{Cognitive Testing}\
All cognitive measures were collected as part of a larger testing battery designed to assess cognitive control and decision making prior to a multi-week cognitive training program. We selected four measures based on (1) their statistical independence from one another and (2) their utility in predicting functional outcomes in health and disease. Specifically, measures included the continuous performance test number correct \cite{kurtz2001comparison}, color shape median switch cost (response times) \cite{miyake2004inner}, median Stroop effect (response times) \cite{stroop1935studies}, and a fractal (visuo-spatial) n-back \cite{ehlis2008reduced,green2005muscarinic,owen2005n}. All tasks were administered to participants seated at a desktop computer approximately two feet from the monitor. All stimuli were presented on a black background. 

\paragraph*{Continuous performance test.}
The continuous performance test consisted of 360 trials of sequentially presented configurations of 2 to 7 red bars representing a digital display.  Each trial was 1s in length with a fixed 700ms intertrial interval before the presentation of the next stimulus. One third of the stimuli (120 out of 360) formed numbers or letters. Participants were instructed to respond with a right index finger press of the spacebar on a standard keyboard when a number or letter was observed. No response was required for non-target trials. Responses made less than 100ms after the presentation of a stimulus were counted as incorrect due to inability to know if such responses were a reaction to the previous trial or impulsive response to the current trial \cite{kurtz2001comparison}.

\paragraph*{Color shape task.}
	The color shape task consisted of two blocks of 48 trials (96 total) of sequentially presented figures with either a circle or triangle in the center flanked by either a red or green square surrounding the shape.  Trial cues were presented 150ms before the stimulus, and both the cue and the stimulus remained on the screen until the participant responded. The cue ``C" indicated that the participant should respond to the color (the ``z'' key with left index finger = green, the backslash key with right index finger = red), whereas an ``S" indicated that the participant should respond to the shape (the ``z'' key with left index finger = circle, the backslash key with right index finger = triangle). Stimuli were separated by a 350ms blank screen and remained on screen until the participant responded \cite{miyake2004inner}. 

\paragraph*{Stroop task.}
The Stroop task consisted of 96 words printed on the screen sequentially in either red, green, or blue. Trials lasted until either the participant responded or 3.5 seconds elapsed and were separated by a fixed 100ms intertrial interval with a blank screen. In congruent trials, the printed word matched the color of the word. In incongruent trials, the color did not match the printed word. Participants were asked to respond only to the color of the word by pressing a colored button with one of the first three fingers on their right hand that matched the word color on each trial. 

\paragraph*{Visual-spatial N-back.}
During the n-back, participants were instructed to remember the location of a stimulus, a grey circle (approximately 5 cm in diameter), as it appeared randomly in 8 possible locations around the perimeter of a computer screen. The stimulus appeared for 200 ms, followed by an interstimulus interval (ISI) of 2800 ms. A cross hair remained visible during the stimulus presentation to cue participants to look at the center of the screen so that all stimuli appearing around the perimeter of the screen could be seen clearly.  The n-back task includes 4 conditions of 50 trials each of varying difficulty levels (200 trials total): the 0-back, 1-back, 2-back, and 3-back. During the 0-back, participants were asked to respond if the stimulus appeared in the upper left corner of the screen. During the 1-back condition, participants respond if the image is identical to the one preceding it.  In the 2-back condition, they respond if the stimulus is identical to the one two trials before.  In the 3-back condition they respond if the stimulus is identical to the one three trials before. Participants were asked to respond only to targets (30\% of stimuli; 60 trials) by pressing the spacebar \cite{ehlis2008reduced,green2005muscarinic,owen2005n}. The primary outcome is the total number of true positives (correctly identifying a target) summed across all conditions.

\paragraph*{Behavioral Results.}
See Table 1 for the mean and standard deviation of performance for the four cognitive measures across subjects. See Figure 7 for the inter-task correlations for the four measures.
\begin{table}[ht]
\caption{Sample performance on the four cognitive control tasks} 
\begin{centering}
\begin{tabular}{c c c c} 
\hline\hline 
CPT(\# correct) & Switch cost(ms) &  Stroop(ms) & N-Back(\#correct) \\ [0.5ex] 
\hline 
110.7 (9.6) & 322.5 (253.5) & 76.9 (89.5) & 52.4 (4.9) \\  [1ex]
\hline 
\end{tabular}
\label{table:nonlin} 
\end{centering}
Cognitive task performance is reported as mean (standard deviation) across all 125 participants. CPT maximum \# correct = 120. N-Back maximum number correct = 60.
\end{table}

\begin{figure}[h!]
		\centerline{\includegraphics[width=3.5in]{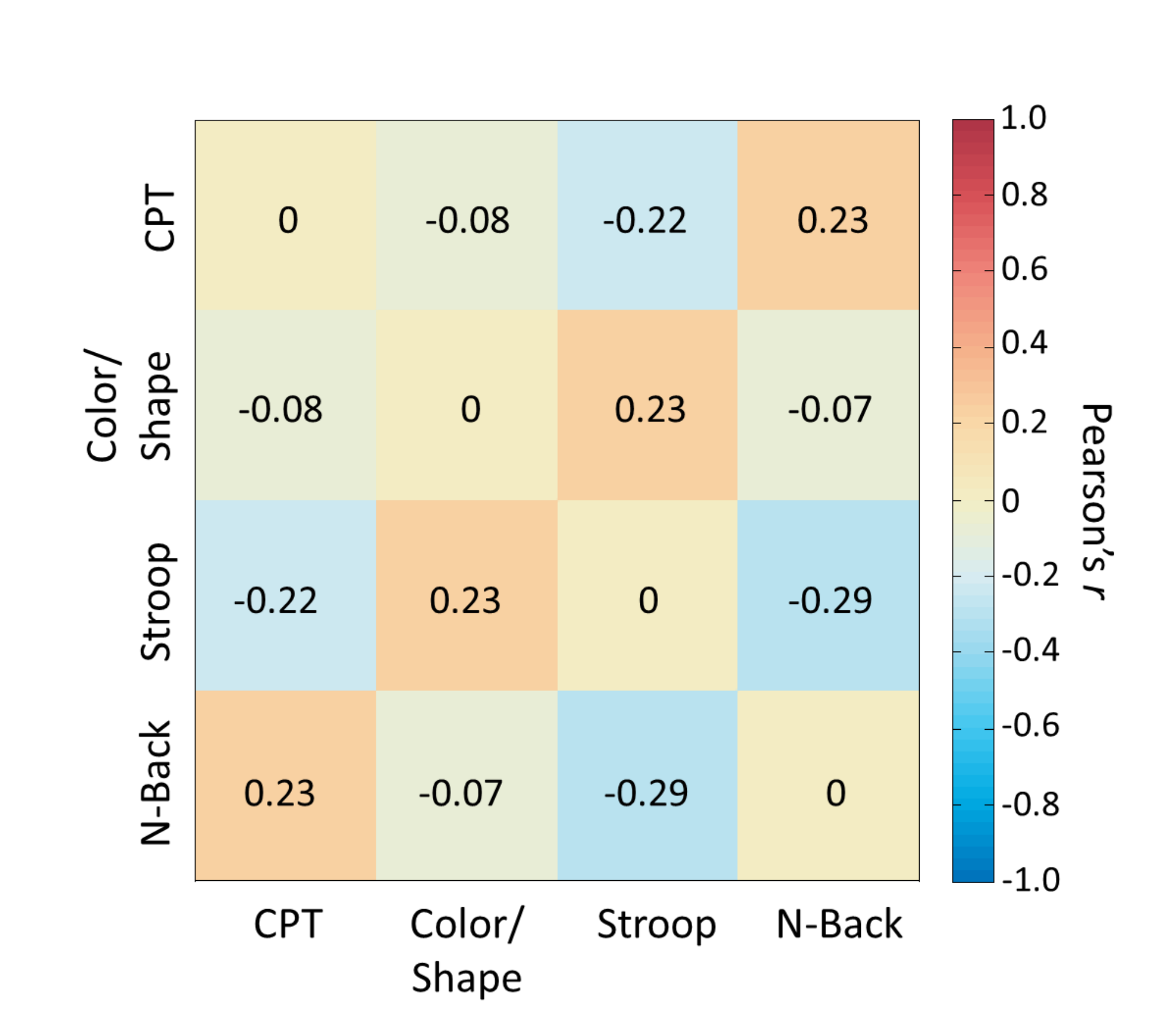}}
		\caption{\textbf{Correlations among cognitive measures.} Numbers in each cell represent the Pearson's correlation coefficient for each pairwise relationship between behavioral measures. The diagonal values were set to zero in this illustration. CPT = continuous performance test number correct, color/shape = color/shape task median switch cost, SSRT = stop signal reaction time, Stroop = median Stroop effect, n-back = total true positives summed over the 1-back, 2-back, and 3-back.} \label{cognitivemeasures}
	\end{figure}

\subsection*{Network Control Theory}\

		We follow a previous application of network control theory in diffusion weighted imaging data \cite{Gu2015}. While we briefly describe the mathematical approach here, we also point the reader to Ref.~ \cite{Gu2015} for a more comprehensive description, and to Refs.~ \cite{Liu2011,Ruths2014,Pasqualetti2014} for additional mathematical details.
		
		Our ability to understand neural systems is fundamentally related to our theoretical ability to control them \cite{Schiff2012}. Network control theory is a branch of traditional control theory in engineering that examines how to control a system based on the links between its components; here, \emph{control} means perturbing a system to reach a desired state. To apply a network control perspective, we require (i) knowledge of the network connectivity linking system components, and (ii) knowledge regarding how system components act, i.e., their \emph{dynamics}. In contrast to traditional graph theory, network control theory offers mechanistic predictors of network dynamics. The use of mechanistic models offers the possibility for us to move from descriptive approaches in the human connectome toward an understanding of the structural and functional bases of cognition \cite{Medaglia2015}.
		
		Mathematically, we can study the controllability of a networked system by defining a network represented by the graph $\mc G = (\mc V, \mc E)$, where $\mc V$ and $\mc E$ are the vertex and edge sets, respectively. Let $a_{ij}$ be the weight associated with the edge $(i,j) \in \mc E$, and define the \emph{weighted adjacency matrix} of $\mc G$ as $A = [a_{ij}]$, where $a_{ij} = 0$ whenever $(i,j) \not\in \mc E$. We associate a real value (\emph{state}) with each node, collect the node states into a vector (\emph{network state}), and define the map $\map{x}{\mathbb{N}_{\ge 0}}{\mathbb{R}^n}$ to describe the evolution (\emph{network dynamics}) of the network state over time. Given the network and node dynamics, we can use network control theory to quantitatively examine how the network structure predicts the types of control that nodes can exert.
		
\subsection*{Network Control Theory Applied to Neuroimaging}\

We begin with an analogous approach to prior work \cite{Gu2015}. We define structural brain networks by subdividing the entire brain into anatomically distinct brain areas (network nodes) in a commonly used anatomical atlas \cite{Hagmann2008}. Consistent with prior work \cite{Bassett2011,Hermundstad2013,Hermundstad2014,Gu2015}, we connect nodes by the number of white matter streamlines identified by a commonly used deterministic tractography algorithm (see \cite{Cieslak2014} and above description). This procedure results in sparse, weighted, undirected structural brain networks for each subject (N = 125). The definition of structural brain networks based on tractography data in humans follows from our primary hypothesis that cognitive control processes are in part determined by the structural organization of the brain's white matter tracts.
		
To define the dynamics of neural processes, we draw on prior models linking structural brain networks to resting state functional dynamics \cite{honey2009predicting,honey2010can,abdelnour2014network}. Although neural activity evolves through neural circuits as a collection of \emph{nonlinear} dynamic processes, these prior studies have demonstrated that a significant amount of variance in neural dynamics as measured by fMRI can be predicted from simplified \emph{linear} models \cite{honey2009predicting,Galan2008}. Based on this literature, we employ a simplified noise-free linear discrete-time and time-invariant network model: 
		
		\begin{equation}\label{eq: linear network}
			\mathbf{x} (t+1) = \mathbf{A} \mathbf{x}(t) + \mathbf{B}_{\mc K} \mathbf{u}_{\mc K} (t),
		\end{equation}
		
		where $\map{\mathbf{x}}{\real_{\ge 0}}{\real^N}$ describes the state (e.g., a measure of the electrical charge, oxygen level, or firing rate) of brain regions over time, and $\mathbf{A} \in \real^{N \times N}$ is a symmetric and weighted adjacency matrix. In this case, we construct a weighted adjacency matrix whose elements indicate the number of white matter streamlines connecting two different brain regions -- denoted here as $i$ and $j$ -- and we stabilize this matrix by dividing by the mean edge weight. The diagonal elements of the matrix $\mathbf{A}$ satisfy $A_{ii}=0$. The input matrix $\mathbf{B}_{\mc K}$ identifies the control points $\mc K$ in the brain, where $\mc K = \{k_1, \dots, k_m \}$ and \begin{align}\label{eq: B}
			B_{\mc K} =
			\begin{bmatrix}
				e_{k_1} & \cdots & e_{k_m}
			\end{bmatrix},
		\end{align}
		and $e_i$ denotes the $i$-th canonical vector of dimension $N$. The input $\map{\mathbf{u}_{\mc K}}{\real_{\ge  0}}{\real^m}$ denotes the control strategy.

\subsection*{Network Controllability}\

To study the ability of a certain brain region to influence other regions in arbitrary ways we adopt the control theoretic notion of \emph{controllability}. Controllability of a dynamical system refers to the possibility of driving the state of a dynamical system to a specific target state by means of an external control input \cite{REK-YCH-SKN:63}. Classic results in control theory ensure that controllability of the network \eqref{eq: linear network} from the set of network nodes $\mc K$ is equivalent to the controllability Gramian $\mathbf{W}_{\mc K}$ being invertible, where \begin{equation}
			\mathbf{W}_{\mathcal{K}} = \sum_{\tau =0}^{\infty}\mathbf{A}^\tau
			\mathbf{B}_{\mathcal{K}}\mathbf{B}_{\mathcal{K}}^\transpose \mathbf{A}^\tau .
		\end{equation}
		We utilize this framework to choose control nodes one at a time, and thus the input matrix $B$ in fact reduces to a one-dimensional vector.
		
		Besides ensuring controllability, the eigenvalues of the controllability Gramian are a quantitative measure of the magnitude of the control input that drives a network to a desired target state \cite{TK:80}, and the structure of the Gramian itself provides systematic guidelines for the selection of control areas that can theoretically optimize cognitive functions. While the magnitude of the control input may not be the unique feature to take into account when controlling brain dynamics \cite{kumar2014input}, it allows us to better understand the relationship between the structural organization of the brain and its dynamics, and opens the door to the development of novel diagnostics. Here, this allows us to isolate the control role of each region separately in the context of the individual's network, and then to associate variability in this property at each node with cognition.

\subsection*{Network Controllability Statistics}\
In this work, we study two types of controllability: modal controllability, and boundary controllability. Intuitively, modal controllability refers to the ability of a node to control each evolutionary mode of a dynamical network \cite{AMAH-AHN:89}, and can be used to identify network configurations that are least controllable. Modal controllability is computed from the eigenvector matrix $V = [v_{ij}]$ of the network adjacency matrix $\mathbf{A}$. By extension from the PBH test \cite{TK:80}, if the entry $v_{ij}$ is small, then the $j$-th mode is poorly controllable from node $i$. Following \cite{FP-SZ-FB:13q}, we define $\phi_i =\sum_{j =1}^N (1 - \lambda_j^2 (A)) v_{ij}^2$ as a scaled measure of the controllability of all $N$ modes $\lambda_1 (A),\dots, \lambda_N(A)$ from the brain region $i$. Regions with high modal controllability are able to control all the dynamic modes of the network, and hence to drive the dynamics towards hard-to-reach configurations. Modal control values were assigned a ranked value to linearize their distribution \cite{gu2015controllability}.

To complement modal controllability, we also study \emph{boundary controllability}, a metric developed in network control theory to quantify the role of a network node in controlling dynamics between modules in hierarchical networks. Boundary controllability identifies brain areas that can steer the system into states where different cognitive systems are either coupled or decoupled. Here, we apply a similar approach to that taken in \cite{Gu2015} to quantify boundary controllability in our diffusion tractography networks and associate controllability variability with cognitive performance.
				 
As boundary controllability describes a node's role in controlling dynamics across modular network architecture, an initial identification of brain modules is required \cite{Gu2015}. To quantify the initial partition for the structural brain networks, we define the first level of the network empirically by maximizing the modularity quality function \cite{Newman2006} using a Louvain-like \cite{Blondel2008} locally greedy algorithm \cite{Jutla2011}. Our choice is based on extensive recent literature demonstrating that the brain is composed of many subnetworks (not just 2) \cite{Meunier2010,Bassett2015}, which can be extracted using modularity maximization approaches \cite{Meunier2009,Chen2008,Bassett2013,Bassett2011b}, and which correspond to sets of brain areas performing related functions \cite{Power2012,Chen2008,Bassett2015}.
		
The modularity quality function provides an estimate of the quality of a hard partition of the $N \times N$ adjacency matrix $\mathbf{A}$ into network communities (whereby each brain region is assigned to exactly one network community) \cite{NG2004,markfast,Newman2006,Porter2009,Fortunato2010}
		\begin{equation}\label{one}
			Q_{0} = \sum_{ij} [A_{ij} - \gamma P_{ij}] \delta(g_{i},g_{j})\,,
		\end{equation}
where brain region $i$ is assigned to community $g_{i}$, brain region $j$ is assigned to community $g_{j}$, $\delta(g_{i},g_{j})=1$ if $g_{i} = g_{j}$ and it equals $0$ otherwise, $\gamma$ is a structural resolution parameter, and $P_{ij}$ is the expected weight of the edge connecting node $i$ and node $j$ under a specified null model. Maximization of $Q_{0}$ yields a hard partition of a network into communities such that the total edge weight inside of communities is as large as possible (relative to the null model and subject to the limitations of the employed computational heuristics, as optimizing $Q_{0}$ is NP-hard \cite{Porter2009,Fortunato2010,Brandes2008}).
		
Because the modularity quality function has many near-degeneracies, it is important to perform the optimization algorithm multiple times \cite{Good2010}. We perform 100 optimizations of the Louvain-like locally greedy algorithm \cite{Jutla2011} applied to the average structural adjacency matrix. To distill a single representative partition, we create a consensus partition from these 100 optimizations based on statistical comparison to an appropriate null model \cite{Bassett2013}.
		
		Importantly, we choose a value for the structural resolution parameter $\gamma$ that produces a robust consensus partition over individuals. The choice $\gamma  = 1$ is a common default decision, but it is important to consider multiple values of $\gamma$ to examine the multiscale community structure of the \textbf{A} matrix \cite{rb2006,Porter2009,Onnela2011}. Previous work has demonstrated that in some networks, a structural resolution parameter value that accurately captures the true community structure can be identified by the $\gamma$ value at which the 100 optimizations produce similar partitions \cite{Bassett2013}. To quantitatively estimate similarity in partitions, we adopt the $z$-score of the Rand coefficient \cite{Traud2010}. For each pair of partitions $\alpha$ and $\beta$, we calculate the Rand $z$-score in terms of the total number of pairs of nodes in the network $M$, the number of pairs $M_{\alpha}$ that are in the same community in partition $\alpha$, the number of pairs $M_{\beta}$ that are in the same community in partition $\beta$, and the number of pairs of nodes $w_{\alpha \beta}$ that are assigned to the same community both in partition $\alpha$ and in partition $\beta$. The $z$-score of the Rand coefficient comparing these two partitions is
		\begin{equation}
			z_{\alpha\beta} = \frac{1}{\sigma_{w_{\alpha \beta}}} w_{\alpha \beta}-\frac{M_{\alpha}M_{\beta}}{M}\,,
		\end{equation}

where $\sigma_{w_{\alpha \beta}}$ is the standard deviation of $w_{\alpha \beta}$. Let the \emph{mean partition similarity} denote the mean value of $z_{\alpha \beta}$ over all possible partition pairs for $\alpha \neq \beta$. Let the \emph{variance of the partition similarity} denote the variance of $z_{\alpha \beta}$ over all possible partition pairs for $\alpha \neq \beta$.
		
Empirically, we calculated a group adjacency matrix by averaging the structural \textbf{A} matrices of all subjects. We optimized the modularity quality function 100 times and we computed the mean and variance of the partition similarity for a range of $\gamma$ values. We observed that the mean partition similarity was high and the variance of the partition similarity was low for a value of $\gamma$ at $1.7$, which is within the range of stable partitions found in our prior analyses in diffusion spectrum imaging data \cite{Gu2015}. We therefore used the consens partition at $\gamma = 1.7$ for the remainder of the analysis in this study. To examine the fit of the group average consensus partition to each individual, we conducted a permutation test for each individual where the community assignments were randomly permuted 10,000 times per individual. Then, we computed the mean and variability of the $z$-score of the Rand coefficient across all subjects and found that the subject similarity to the consensus partition was superior to randomly permuted assignments for all subjects (mean $z$-Rand score = 55.4, standard deviation = 3.4 over all permutations for all subjects). 
		
\paragraph*{Boundary Point Criteria}  The first modification concerns the definitions of the first level subnetworks for which we compute a two-partition based on the Fiedler eigenvector. Consistent with our prior work \cite{gu2015controllability,FP-SZ-FB:13q} within the initial community partition we compute the Fiedler eigenvector of the adjacency matrix to create first level subnetworks defined by a two-partition.  After calculting the two community partition, we must identify ``boundary points'', which are nodes that contain connections to both communities. Following \cite{Gu2015}, we set a threshold ratio $\rho$ to identify boundary points. Considering the adaptivity to the local measure, we set a threshold ratio $\rho$ instead of a global threshold value. In detail, for a network $G =(V, E)$ with partition $P = (V_1, \cdots, V_n)$, a node $i\in V_k$ is called a boundary node if

				\begin{equation}
				\sum_{l\neq k}a_{kl} \geq \rho\cdot\max(A)
				\end{equation}
				
where $A$ is the adjacency matrix. Here, $\max(A)$ can be replaced with other statistics and $\rho$ needs to be chosen carefully. If $\rho$ is too small, there will be no effect of differences in edge weights within the subnetwork and the algorithm tends to add the total subnetwork as the set of boundary points. If $\rho$ is too large, there will be only a few points recognized as the boundary points.
				
\begin{figure}[h]
				 \centerline{\includegraphics[width=3.5in]{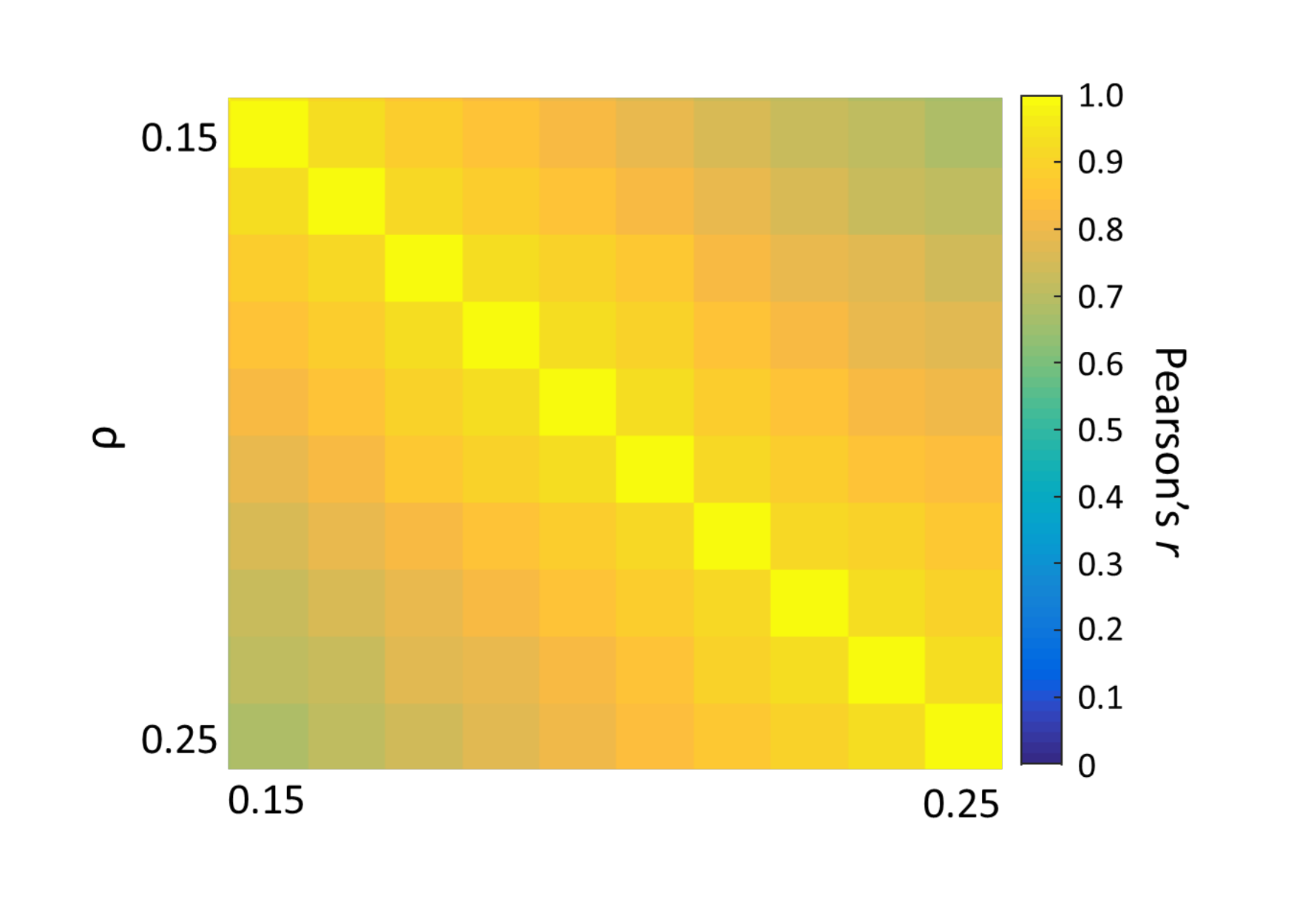}}
				\caption{textbf{Effect of Boundary Point Criteria Threshold}  Color indicates Pearson correlation coefficient, $r$, between the vectors of boundary controllability values estimated for pairs of $\rho$ values in the range $0.15-0.25$ in increments of $0.01$. \label{rho}}
				 \end{figure}

In the results described in the main manuscript, we computed boundary controllability ranked over a range of $\rho = 0.15$ to $0.25$ in increments of $0.01$. The boundary control values are highly similar across choices of $\rho$ (See Fig.~\ref{rho}; minimum Pearson correlation approximately $0.74$, corresponding to a $p = 0$, indicating that our results are robust to small variation in the boundary point criteria threshold. To use a stable measure of boundary controllability for association with cognitive variability, we compute boundary controllability for each node at each value of $\rho$ and take the average value of boundary controllability over this range.
		
		\textbf{Final Algorithm}
		The final algorithm used in the calculation of boundary controllability in this paper can be summarized as follows. We begin with the application of a community detection method to the brain network to extract a partition of brain regions into network communities. We then recursively apply a Fiedler bipartition to add boundary nodes within communities, with the goal of improving the local controllability of the network. At each stage of the algorithm, we define the boundary nodes of the network as the nodes that maintain edges to nodes in other communities. Algorithmically, we can write:
		
		\begin{algorithm}[H]
		 \SetAlgoLined
		 \KwData{Network $G = (V, E)$ with adjacency matrix $A = (a_{ij})$, Number of control nodes $m$, threshold ratio $\rho$;}
		 \KwResult{Control Nodes Index Set $\mathcal{K}$; }
		 Define an empty set of control nodes $\mathcal{K} = \emptyset$\;
		 Initialize the partition $\mathcal{P}$ with the result of a community detection algorithm and initialize the boundary nodes set $\mathbf{B} = \emptyset$\;
		 Add the boundary points of the initial partition\;
		 \While{$|\mathcal{K}| < m$}{
		  Select least controllable community $l = \arg\min\{\lambda_{min}(W_{i,\infty}), i = 1, ..., |\mathcal{P}|\}$\;
		  Compute Fiedler two-partition $P_f$ of $l$-th community\;
		  Compute boundary nodes $B_f$ of $P_f$ with the given threshold ratio $\rho$\;
		  Update partition $\mathcal{P}$ with $P_f$\;
		  Update control nodes with boundary nodes $\mathcal{K} = \mathcal{K}\cup B_f$\;
		  }
		   \Return $\mathcal{K}$.
		 \caption{Algorithm for the Selection of Boundary Control Nodes}\label{alg1}
		\end{algorithm}

\subsection*{Associating Controllability with Cognition}\

		To associate network controllability with measured cognitive performance, we computed the partial correlation between each cognitive variable and ranked controllability values controlling for age, sex, and DTI-scan total signal to noise values for each participant. Partial correlation coefficients were Fisher $r$-to-$z$ transformed for visualization in the final maps. For each map, we tested for significant associations between the strength of controllability and geometric distance between regions. For each node pair in the adjacency matrix $A_{ij}$, we computed the Pearson's correlation between the mean absolute value for the $z$-transformed partial correlation coefficients and the Euclidean distance between region centers (two-sided test). The distance for region centers were transformed  by empirically calculating the cumulative density function of observed distances and transforming this distribution to normal. Within the maps, nodes surpassing a Fisher's $r$ value of +/- 2.0 were interpreted in the main manuscript.
		
\subsection*{Code Availability}
	Code used to calculate the controllability statistics described here can be found at: http://commdetect.weebly.com. 

\subsection*{Data Availability}
	Data are available on request to the authors contingent upon IRB authorization.

\section*{Acknowledgements}
			JDM acknowledges support from the Office of the Director at the National Institutes of Health through grant number 1-DP5-OD-021352-01.  DSB acknowledges support from the John D. and Catherine T. MacArthur Foundation, the Army Research Laboratory and the Army Research Office through contract numbers W911NF-10-2-0022 and W911NF-14-1-0679, the National Institute of Mental Health (2-R01-DC-009209-11), the National Institute of Child Health and Human Development (1R01HD086888-01), the Office of Naval Research, and the National Science Foundation (\#BCS-1441502, \#BCS-1430087, and \#PHY-1554488). SG acknowledges support from the Applied Mathematics and Computational Science Graduate Program at the University of Pennsylvania. FP acknowledges support from the National Science Foundation award \#BCS-1430279. CL acknowledges support from the National Cancer Institute (\#R35-CA197461 and \#R01-CA170297).  The content is solely the responsibility of the authors and does not necessarily represent the official views of any of the funding agencies.

		\newpage
		
\bibliography{JDMReferences}

\begin{thebibliography}{100}

\bibitem{Normandeau1998}
S.~Normandeau, F.~Guay, {\it Journal of Educational Psychology\/} {\bf 90}, 111
  (1998).

\bibitem{Diamond2007}
A.~Diamond, W.~S. Barnett, J.~Thomas, S.~Munro, {\it Science (New York, NY)\/}
  {\bf 318}, 1387 (2007).

\bibitem{Cole2013}
M.~W. Cole, {\it et~al.\/}, {\it Nat Neurosci\/} {\bf 16}, 1348 (2013).

\bibitem{Power2013}
J.~D. Power, B.~L. Schlaggar, C.~N. Lessov-Schlaggar, S.~E. Petersen, {\it
  Neuron\/} {\bf 79}, 798 (2013).

\bibitem{Voytek2015}
B.~Voytek, R.~T. Knight, {\it Biological psychiatry\/} {\bf 77}, 1089 (2015).

\bibitem{Medaglia2015}
J.~D. Medaglia, M.~E. Lynall, D.~S. Bassett, {\it J Cogn Neurosci\/} {\bf 27},
  1471 (2015).

\bibitem{Cocchi2013}
L.~Cocchi, A.~Zalesky, A.~Fornito, J.~B. Mattingley, {\it Trends in cognitive
  sciences\/} {\bf 17}, 493 (2013).

\bibitem{Gu2015}
S.~Gu, {\it et~al.\/}, {\it Nature Communications\/} {\bf 6}, 8414 (2015).

\bibitem{Mattar2015}
M.~G. Mattar, M.~W. Cole, S.~L. Thompson-Schill, D.~S. Bassett, {\it PLoS
  computational biology\/} {\bf 11} (2015).

\bibitem{Braun2015}
U.~Braun, {\it et~al.\/}, {\it Proceedings of the National Academy of
  Sciences\/} {\bf 112}, 11678 (2015x).

\bibitem{Bassett2015}
D.~S. Bassett, M.~Yang, N.~F. Wymbs, S.~T. Grafton, {\it Nat Neurosci\/} {\bf
  18}, 744 (2015).

\bibitem{Ruths2014}
J.~Ruths, D.~Ruths, {\it Science\/} {\bf 343}, 1373 (2014).

\bibitem{Pasqualetti2014}
F.~Pasqualetti, S.~Zampieri, F.~Bullo, {\it Control of Network Systems, IEEE
  Transactions on\/} {\bf 1}, 40 (2014).

\bibitem{betzel2016optimally}
R.~F. Betzel, S.~Gu, J.~D. Medaglia, F.~Pasqualetti, D.~S. Bassett, {\it arXiv
  preprint arXiv:1603.05261\/}  (2016).

\bibitem{miyake2000unity}
A.~Miyake, {\it et~al.\/}, {\it Cognitive psychology\/} {\bf 41}, 49 (2000).

\bibitem{kurtz2001comparison}
M.~M. Kurtz, J.~D. Ragland, W.~Bilker, R.~C. Gur, R.~E. Gur, {\it Schizophrenia
  research\/} {\bf 48}, 307 (2001).

\bibitem{miyake2004inner}
A.~Miyake, M.~J. Emerson, F.~Padilla, J.-c. Ahn, {\it Acta psychologica\/} {\bf
  115}, 123 (2004).

\bibitem{stroop1935studies}
J.~R. Stroop, {\it Journal of experimental psychology\/} {\bf 18}, 643 (1935).

\bibitem{green2005muscarinic}
A.~Green, {\it et~al.\/}, {\it Pharmacology Biochemistry and Behavior\/} {\bf
  81}, 575 (2005).

\bibitem{ogg2008neural}
R.~J. Ogg, {\it et~al.\/}, {\it Magnetic resonance imaging\/} {\bf 26}, 504
  (2008).

\bibitem{leech2014role}
R.~Leech, D.~J. Sharp, {\it Brain\/} {\bf 137}, 12 (2014).

\bibitem{sali2016spontaneous}
A.~W. Sali, S.~M. Courtney, S.~Yantis, {\it The Journal of Neuroscience\/} {\bf
  36}, 445 (2016).

\bibitem{botvinick2004conflict}
M.~M. Botvinick, J.~D. Cohen, C.~S. Carter, {\it Trends in cognitive
  sciences\/} {\bf 8}, 539 (2004).

\bibitem{nagahama1999transient}
Y.~Nagahama, {\it et~al.\/}, {\it Neuroimage\/} {\bf 10}, 193 (1999).

\bibitem{smith2006task}
A.~B. Smith, E.~Taylor, M.~Brammer, B.~Toone, K.~Rubia, {\it American Journal
  of Psychiatry\/}  (2006).

\bibitem{holland1999amygdala}
P.~C. Holland, M.~Gallagher, {\it Trends in cognitive sciences\/} {\bf 3}, 65
  (1999).

\bibitem{shanmugan2016common}
S.~Shanmugan, {\it et~al.\/}, {\it American Journal of Psychiatry\/}  (2016).

\bibitem{isoda2007switching}
M.~Isoda, O.~Hikosaka, {\it Nature neuroscience\/} {\bf 10}, 240 (2007).

\bibitem{hikosaka2010switching}
O.~Hikosaka, M.~Isoda, {\it Trends in cognitive sciences\/} {\bf 14}, 154
  (2010).

\bibitem{floresco2006dissociable}
S.~B. Floresco, S.~Ghods-Sharifi, C.~Vexelman, O.~Magyar, {\it The Journal of
  neuroscience\/} {\bf 26}, 2449 (2006).

\bibitem{garner2015training}
K.~Garner, P.~E. Dux, {\it Proceedings of the National Academy of Sciences\/}
  {\bf 112}, 14372 (2015).

\bibitem{aron2007converging}
A.~R. Aron, {\it et~al.\/}, {\it The Journal of Neuroscience\/} {\bf 27}, 11860
  (2007).

\bibitem{parent1995functional}
A.~Parent, L.-N. Hazrati, {\it Brain Research Reviews\/} {\bf 20}, 91 (1995).

\bibitem{cools2001mechanisms}
R.~Cools, R.~A. Barker, B.~J. Sahakian, T.~W. Robbins, {\it Brain\/} {\bf 124},
  2503 (2001).

\bibitem{derakshan2009effects}
N.~Derakshan, S.~Smyth, M.~W. Eysenck, {\it Psychonomic Bulletin \& Review\/}
  {\bf 16}, 1112 (2009).

\bibitem{gross2004developmental}
C.~Gross, R.~Hen, {\it Nature Reviews Neuroscience\/} {\bf 5}, 545 (2004).

\bibitem{puce1996differential}
A.~Puce, T.~Allison, M.~Asgari, J.~C. Gore, G.~McCarthy, {\it The Journal of
  Neuroscience\/} {\bf 16}, 5205 (1996).

\bibitem{ludersdorfer2015accessing}
P.~Ludersdorfer, M.~Kronbichler, H.~Wimmer, {\it Human brain mapping\/} {\bf
  36}, 1393 (2015).

\bibitem{creem2001defining}
S.~H. Creem, D.~R. Proffitt, {\it Acta psychologica\/} {\bf 107}, 43 (2001).

\bibitem{banich2000fmri}
M.~T. Banich, {\it et~al.\/}, {\it Cognitive Neuroscience, Journal of\/} {\bf
  12}, 988 (2000).

\bibitem{koenigs2009superior}
M.~Koenigs, A.~K. Barbey, B.~R. Postle, J.~Grafman, {\it The Journal of
  Neuroscience\/} {\bf 29}, 14980 (2009).

\bibitem{vandenberghe2012spatial}
R.~Vandenberghe, P.~Molenberghs, C.~R. Gillebert, {\it Neuropsychologia\/} {\bf
  50}, 1092 (2012).

\bibitem{braga2013echoes}
R.~M. Braga, D.~J. Sharp, C.~Leeson, R.~J. Wise, R.~Leech, {\it The Journal of
  Neuroscience\/} {\bf 33}, 14031 (2013).

\bibitem{petrides2000role}
M.~Petrides, {\it Executive Control and the Frontal Lobe: Current Issues\/}
  (Springer, 2000), pp. 44--54.

\bibitem{Power2011}
J.~D. Power, {\it et~al.\/}, {\it Neuron\/} {\bf 72}, 665 (2011).

\bibitem{brandt2016selective}
K.~R. Brandt, M.~W. Eysenck, M.~K. Nielsen, T.~J. von Oertzen, {\it Brain and
  cognition\/} {\bf 104}, 82 (2016).

\bibitem{olson2007enigmatic}
I.~R. Olson, A.~Plotzker, Y.~Ezzyat, {\it Brain\/} {\bf 130}, 1718 (2007).

\bibitem{bechara2000emotion}
A.~Bechara, H.~Damasio, A.~R. Damasio, {\it Cerebral cortex\/} {\bf 10}, 295
  (2000).

\bibitem{berlin2005borderline}
H.~A. Berlin, E.~T. Rolls, S.~D. Iversen, {\it American journal of
  psychiatry\/} {\bf 162}, 2360 (2005).

\bibitem{dziobek2011neuronal}
I.~Dziobek, {\it et~al.\/}, {\it Neuroimage\/} {\bf 57}, 539 (2011).

\bibitem{Bullmore2009}
E.~Bullmore, O.~Sporns, {\it Nature Review Neuroscience\/} {\bf 10}, 186
  (2009).

\bibitem{Rubinov2010}
M.~Rubinov, O.~Sporns, {\it NeuroImage\/} {\bf 52}, 1059 (2010).

\bibitem{Bullmore2011}
E.~T. Bullmore, D.~S. Bassett, {\it Annual Reviews of Clinical Psychology\/}
  {\bf 7}, 113 (2011).

\bibitem{Sepulcre2014}
J.~Sepulcre, {\it Neuroscience Letters\/} {\bf 567}, 68  (2014).

\bibitem{Hellyer2014}
P.~J. Hellyer, {\it et~al.\/}, {\it Journal of Neuroscience\/} {\bf 34}, 451
  (2014).

\bibitem{wang2008functional}
S.-H. Wang, {\it Brain, behavior and evolution\/} {\bf 72}, 159 (2008).

\bibitem{chittka2009speed}
L.~Chittka, P.~Skorupski, N.~E. Raine, {\it Trends in Ecology \& Evolution\/}
  {\bf 24}, 400 (2009).

\bibitem{Pinker1999mind}
S.~Pinker, {\it Annals of the New York Academy of Sciences\/} {\bf 882}, 119
  (1999).

\bibitem{cohen2007should}
J.~D. Cohen, S.~M. McClure, J.~Y. Angela, {\it Philosophical Transactions of
  the Royal Society of London B: Biological Sciences\/} {\bf 362}, 933 (2007).

\bibitem{botvinick2014computational}
M.~M. Botvinick, J.~D. Cohen, {\it Cognitive science\/} {\bf 38}, 1249 (2014).

\bibitem{wedeen2008diffusion}
V.~J. Wedeen, {\it et~al.\/}, {\it Neuroimage\/} {\bf 41}, 1267 (2008).

\bibitem{Miyake2001}
A.~Miyake, N.~P. Friedman, D.~A. Rettinger, P.~Shah, M.~Hegarty, {\it Journal
  of experimental psychology: General\/} {\bf 130}, 621 (2001).

\bibitem{Motter2015}
A.~E. Motter, {\it Chaos: An Interdisciplinary Journal of Nonlinear Science\/}
  {\bf 25}, 097621 (2015).

\bibitem{Kraus2007white}
M.~F. Kraus, {\it et~al.\/}, {\it Brain\/} {\bf 130}, 2508 (2007).

\bibitem{Roalf2016}
D.~R. Roalf, {\it et~al.\/}, {\it NeuroImage\/} {\bf 125}, 903 (2016).

\bibitem{gu2015controllability}
S.~Gu, {\it et~al.\/}, {\it Nature communications\/} {\bf 6} (2015).

\bibitem{yeh2011estimation}
F.-C. Yeh, V.~J. Wedeen, W.-Y.~I. Tseng, {\it Neuroimage\/} {\bf 55}, 1054
  (2011).

\bibitem{fischl2012freesurfer}
B.~Fischl, {\it Neuroimage\/} {\bf 62}, 774 (2012).

\bibitem{cammoun2012mapping}
L.~Cammoun, {\it et~al.\/}, {\it Journal of neuroscience methods\/} {\bf 203},
  386 (2012).

\bibitem{cieslak2014local}
M.~Cieslak, S.~Grafton, {\it Brain imaging and behavior\/} {\bf 8}, 292 (2014).

\bibitem{ehlis2008reduced}
A.-C. Ehlis, C.~G. B{\"a}hne, C.~P. Jacob, M.~J. Herrmann, A.~J. Fallgatter,
  {\it Journal of psychiatric research\/} {\bf 42}, 1060 (2008).

\bibitem{owen2005n}
A.~M. Owen, K.~M. McMillan, A.~R. Laird, E.~Bullmore, {\it Human brain
  mapping\/} {\bf 25}, 46 (2005).

\bibitem{Liu2011}
Y.-Y. Liu, J.-J. Slotine, A.-L. Barab{\'a}si, {\it Nature\/} {\bf 473}, 167
  (2011).

\bibitem{Schiff2012}
S.~J. Schiff, {\it Neural Control Engineering\/} (The MIT Press, 2012).

\bibitem{Hagmann2008}
P.~Hagmann, {\it et~al.\/}, {\it PLoS Biology\/} {\bf 6}, e159 (2008).

\bibitem{Bassett2011}
D.~S. Bassett, J.~A. Brown, V.~Deshpande, J.~M. Carlson, S.~T. Grafton, {\it
  Neuroimage\/} {\bf 54}, 1262 (2011).

\bibitem{Hermundstad2013}
A.~M. Hermundstad, {\it et~al.\/}, {\it Proc Natl Acad Sci U S A\/} {\bf 110},
  6169 (2013).

\bibitem{Hermundstad2014}
A.~M. Hermundstad, {\it et~al.\/}, {\it PLoS Comput Biol\/} {\bf 10}, e1003591
  (2014).

\bibitem{Cieslak2014}
M.~Cieslak, S.~T. Grafton, {\it Brain Imaging Behav\/} {\bf 8}, 292 (2014).

\bibitem{honey2009predicting}
C.~Honey, {\it et~al.\/}, {\it Proceedings of the National Academy of
  Sciences\/} {\bf 106}, 2035 (2009).

\bibitem{honey2010can}
C.~J. Honey, J.-P. Thivierge, O.~Sporns, {\it Neuroimage\/} {\bf 52}, 766
  (2010).

\bibitem{abdelnour2014network}
F.~Abdelnour, H.~U. Voss, A.~Raj, {\it Neuroimage\/} {\bf 90}, 335 (2014).

\bibitem{Galan2008}
R.~Fern\'{a}ndez~Gal\'{a}n, {\it PLoS One\/} {\bf 3}, e2148 (2008).

\bibitem{REK-YCH-SKN:63}
R.~E. Kalman, Y.~C. Ho, S.~K. Narendra, {\it Contributions to Differential
  Equations\/} {\bf 1}, 189 (1963).

\bibitem{TK:80}
T.~Kailath, {\it Linear Systems\/} (Prentice-Hall, 1980).

\bibitem{kumar2014input}
G.~Kumar, D.~Menolascino, S.~Ching, {\it arXiv preprint arXiv:1411.5892\/}
  (2014).

\bibitem{AMAH-AHN:89}
A.~M.~A. Hamdan, A.~H. Nayfeh, {\it {AIAA} Journal of Guidance, Control, and
  Dynamics\/} {\bf 12}, 421 (1989).

\bibitem{FP-SZ-FB:13q}
F.~Pasqualetti, S.~Zampieri, F.~Bullo, {\it IEEE Transactions on Control of
  Network Systems\/} {\bf 1}, 40 (2014).

\bibitem{Newman2006}
M.~E. Newman, {\it Proc Natl Acad Sci U S A\/} {\bf 103}, 8577 (2006).

\bibitem{Blondel2008}
V.~D. Blondel, J.-L. Guillaume, R.~Lambiotte, E.~Lefebvre, {\it Journal of
  Statistical Mechanics: Theory and Experiment\/} {\bf 10}, P1000 (2008).

\bibitem{Jutla2011}
I.~S. Jutla, L.~G.~S. Jeub, P.~J. Mucha, A generalized {L}ouvain method for
  community detection implemented in {MATLAB} (2011--2012).

\bibitem{Meunier2010}
D.~Meunier, R.~Lambiotte, E.~T. Bullmore, {\it Front Neurosci\/} {\bf 4}, 200
  (2010).

\bibitem{Meunier2009}
D.~Meunier, S.~Achard, A.~Morcom, E.~Bullmore, {\it Neuroimage\/} {\bf 44}, 715
  (2009).

\bibitem{Chen2008}
Z.~J. Chen, Y.~He, P.~Rosa-Neto, J.~Germann, A.~C. Evans, {\it Cereb Cortex\/}
  {\bf 18}, 2374 (2008).

\bibitem{Bassett2013}
D.~S. Bassett, F.~Siebenhuhner, {\it In Multiscale Analysis and Nonlinear
  Dynamics: From Genes to the Brain\/} (Wiley, 2013), chap. Multiscale network
  organization in the human brain.

\bibitem{Bassett2011b}
D.~S. Bassett, {\it et~al.\/}, {\it Proc Natl Acad Sci U S A\/} {\bf 108}, 7641
  (2011).

\bibitem{Power2012}
J.~D. Power, {\it et~al.\/}, {\it Neuron\/} {\bf 72}, 665 (2011).

\bibitem{NG2004}
M.~E. Newman, M.~Girvan, {\it Phys Rev E\/} {\bf 69}, 026113 (2004).

\bibitem{markfast}
M.~E. Newman, {\it Phys Rev E\/} {\bf 69}, 066133 (2004).

\bibitem{Porter2009}
M.~A. Porter, J.-P. Onnela, P.~J. Mucha, {\it Notices of the American
  Mathematical Society\/} {\bf 56}, 1082 (2009).

\bibitem{Fortunato2010}
S.~Fortunato, {\it Phys Rep\/} {\bf 486}, 75 (2010).

\bibitem{Brandes2008}
U.~Brandes, {\it et~al.\/}, {\it IEEE Trans on Knowl Data Eng\/} {\bf 20}, 172
  (2008).

\bibitem{Good2010}
B.~H. Good, Y.~A. de~Montjoye, A.~Clauset, {\it Phys Rev E\/} {\bf 81}, 046106
  (2010).

\bibitem{rb2006}
J.~Reichardt, S.~Bornholdt, {\it Phys Rev E\/} {\bf 74}, 016110 (2006).

\bibitem{Onnela2011}
J.-P. Onnela, {\it et~al.\/}, {\it {Phys Rev E}\/} {\bf 86}, 036104 (2012).

\bibitem{Traud2010}
A.~L. Traud, E.~D. Kelsic, P.~J. Mucha, M.~A. Porter, {\it SIAM Review\/} {\bf
  53}, 526 (2011).

\end{thebibliography}

\bibliographystyle{Science}
		
	\end{document}